\documentclass[preprint,twocolumn,superscriptaddress,balancelastpage,showpacs,reprint]{revtex4-2}

\usepackage{blindtext}
\usepackage{centernot}
\usepackage{graphicx}
\usepackage{amsmath,bbold}
\usepackage{times}
\usepackage{amssymb}
\usepackage{mathrsfs}
\usepackage{chemarr}
\usepackage{color}
\usepackage{url}
\usepackage{version}
\usepackage[hidelinks]{hyperref}
\usepackage{mwe,tikz}
\usepackage[percent]{overpic}
\usepackage{bm}
\usepackage[export]{adjustbox}
\definecolor{linkcolor}{rgb}{0,0,0.6} 
\usepackage{lipsum}
\usepackage{pdfpages}

\usepackage{lipsum}

\usetikzlibrary{patterns}

\makeatletter
\AtBeginDocument{\let\LS@rot\@undefined}
\makeatother

\begin{document}
  

\title{Unveiling the Phase Diagram and Reaction Paths of the Active Model B\\
with the Deep Minimum Action Method}

\author{Ruben Zakine}
\affiliation{Courant Institute, New York University, 251 Mercer Street, New York, New York 10012, USA}
\affiliation{Chair of Econophysics and Complex Systems, École Polytechnique, 91128 Palaiseau Cedex, France}
\affiliation{LadHyX UMR CNRS 7646, École Polytechnique, 91128 Palaiseau Cedex, France}

\author{Eric Simonnet}
\affiliation{INPHYNI, Universit\'e C\^ote d’Azur et CNRS, 17 rue Julien Laupr\^etre
06200 Nice, France}

\author{Eric Vanden-Eijnden}
\affiliation{Courant Institute, New York University, 251 Mercer Street, New York, New York 10012, USA}

\date{\today}

\begin{abstract}

Nonequilibrium phase transitions are  notably difficult to analyze because their mechanisms depend on the system's dynamics in a complex way due to the lack of time-reversal symmetry. To complicate matters, the system's steady-state distribution is unknown in general. Here, the phase diagram of the active Model B is computed with a deep neural network implementation of the geometric minimum action method (gMAM). This  approach unveils the unconventional reaction paths and nucleation mechanism in dimensions 1, 2 and 3, by which the system switches between the homogeneous and inhomogeneous phases in the binodal region. Our main findings are: (i) the mean time to escape the phase-separated state is (exponentially) extensive in the system size $L$, but it increases \textit{non-monotonically} with $L$ in dimension 1; (ii) the mean time to escape the homogeneous state is always finite, in line with the recent work of Cates and Nardini~\cite{cates_classical_2022}; (iii) at fixed $L$, the active term increases the stability of the homogeneous phase, eventually destroying the phase separation in the binodal for large but finite systems.
Our results are particularly relevant for active matter systems in which the number of constituents hardly goes beyond $10^7$ and  where finite-size effects matter.

\end{abstract}

\maketitle 

\textit{Introduction--} 
Activated processes, pervasive in nature, are intrinsically difficult to probe in simulations since they require the sampling of rare events~\cite{torrie_nonphysical_1977,bolhuis2002,grafke_numerical_2019,rotskoff_active_2022}. When a first-order phase transition (FOPT) occurs, a nucleation event is usually required for the system to reach its stable phase~\cite{binder1987,cates_classical_2022, richard_speck2016,omar_phase_2021}. 
In  equilibrium systems, we can exploit the property of time-reversal symmetry (TRS) and the knowledge of their equilibrium distribution to derive a free energy from which we can infer both the thermodynamic stability of each phase, and the reaction paths that are followed by the system during activation~\cite{kramers_brownian_1940,onsager1953,e_simplified_2007}. However, TRS breakdown in nonequilibrium systems prevents access to the free energy, necessitating comprehension of activated process mechanisms through dynamics rather than unknown steady-state distributions~\cite{freidlinWentzell1998, ludwig1975, maier_stein1992,dykman_large_1994, bertini2015,bouchet_generalisation_2016, grafke_long_2017,woillez2019}. Mapping their phase diagram thus persists as a challenge.

In this letter, we tackle this issue within the active Model B, a natural nonequilibrium extension of Cahn-Hilliard dynamics with a nonlinear growth term~\cite{kardar1986,sun_CKPZ_1989} breaking TRS. This widely studied model has attracted considerable attention in recent years~\cite{wittkowski2014,nardini2017prx,  nardini2017prx, catesHouches2019,obyrne2023}, serving as an effective description, for instance, of active particles undergoing motility-induced phase separation (MIPS)~\cite{cates_motility_induced_2015,solon2018binodal, speck_critical_2022}.
Here, we map the phase diagram of the active Model B and analyze FOPT pathways. Our findings reveal transitions involving nucleation events markedly different from their equilibrium counterparts, shaped by the interplay between noise and nongradient terms in the stochastic system dynamics. Moreover, in large but finite systems, we demonstrate that the active term can reduce the probability of observing phase-separated state nucleation and facilitate the reverse transition from the phase-separated phase to the homogeneous state.
To obtain these results, we compute reaction paths using a geometric Minimum Action Method (gMAM)~\cite{weinan2004,VE2008gMAM,heymann2008prl}
implementation relying on Physics-Informed Neural Networks (PINNs)~\cite{raissi2019,karniadakis2021}.
This neural implementation, known as deep gMAM~\cite{simonnet_computing_2023}, is notable as it can be transferred to study FOPTs in other nonequilibrium systems. 
It also gives access to higher dimensional problems
not accessible by traditional methods.
Additionally, we cross-check some results of the deep gMAM algorithm using the traditional gMAM method as a benchmark.

\begin{figure}
\includegraphics[width=1\columnwidth]{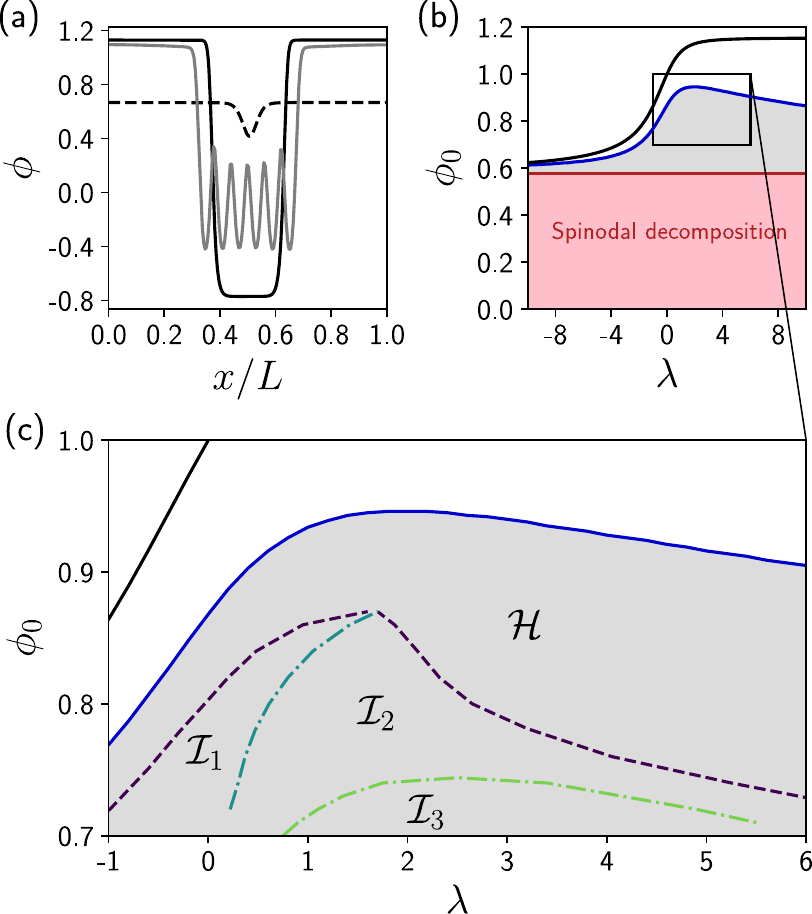}
\caption{(a) Three configurations in $d=1$: The solid line represents the stable inhomogeneous state $\phi_I$, the dashed line indicates the unstable critical state $\phi_{c,1}$, and the grey line depicts a field configuration along the nonequilibrium reaction path from $\phi_I$ to the homogeneous state $\phi_H$ (not shown). Parameters: $\phi_0=0.65$, $\lambda=2$, and $L=120$.
(b) Phase diagram of active Model B in parameter space $(\lambda,\phi_0)$. It shows the binodal (black line) and the spinodal (red line) previously computed in \cite{wittkowski2014}. In finite-size systems, the bistable region does not fully span between the spinodal and the binodal but stops at the blue line (shown here for $L=60$ and $d=1$). Both states $\phi_H$ and $\phi_I$ are stable in the shaded region.
(c) In the bistable region, the purple dashed line marks the FOPT between $\phi_H$ and $\phi_I$. On this line, $V_{\phi_H}(\phi_I)=V_{\phi_I}(\phi_H)$. In region $\mathcal H$, $\phi_H$ is thermodynamically preferred, in regions $\mathcal I_1$, $\mathcal I_2$, $\mathcal I_3$, the inhomogeneous state $\phi_I$ is preferred. The index $q$ in $\mathcal I_q$ indicates the number of bumps along the reaction path from $\phi_I$ to $\phi_H$. Region $\mathcal I_3$ may display asymmetric paths with slightly smaller actions than their symmetric counterparts.}
\label{fig:phase_diagram_amB}
\end{figure}

\textit{Problem setting--} 
The active Model B (AMB) characterizes the stochastic dynamics of a conserved scalar field $\phi(x,t)$, usually interpreted as the local (relative) density of particles or the local composition of a mixture. It can be expressed as the divergence of a noisy flux~\cite{nardini2017prx, tjhung_cluster_2018, cates_classical_2022, wittkowski2014}
\begin{align} 
    \partial_t\phi &=\nabla\cdot(M\nabla \mu+\xi),\label{eq:conserved_dyn}\\
    \mu([\phi],x) &=\frac{\delta \mathcal F[\phi]}{\delta\phi(x)} + \lambda |\nabla\phi(x)|^2,
    \label{eq:chemical_potential}
\end{align}
In this context, $\mathcal F[\phi]$ represents a Ginzburg-Landau free energy, $M$ is the mobility operator, and $\xi$ is a spatio-temporal white noise, a Gaussian process with mean zero and covariance $\langle \xi(x,t)\xi(x',t')\rangle=2\epsilon M\delta(x-x')\delta(t-t')$, where $\epsilon$ controls the amplitude of fluctuations. We investigate Eq.~\eqref{eq:conserved_dyn} in $d=1$ up to $d=3$ dimensions, assuming periodic boundary conditions of the domain $\Omega= [0,L]^d$ with lateral size $L$. We simplify by considering $M=\mathbb 1$ and $\mathcal F[\phi] = \int_\Omega [\frac12\nu(\nabla\phi)^2 +f(\phi)] dx$, where $\nu>0$ and $f(\phi)$ represents a double-well potential. With this choice, there is a region of the phase diagram where a homogeneous state, denoted $\phi_H$, coexists with a phase-separated state (or inhomogeneous state), denoted $\phi_I$ (see Fig.~\ref{fig:phase_diagram_amB}(a)). These states correspond to the two locally stable fixed points of the noiseless version of Eq.~\eqref{eq:conserved_dyn}, namely the solution to 
$\nabla\cdot(M\nabla \mu) =0$ with a prescribed value of the 
spatial average $\phi_0$ of $\phi$ in the domain.

When $\lambda=0$, $\mu$ is the functional derivative of the free energy $\mathcal F[\phi]$, and the dynamics is in detailed balance with respect to the Gibbs-Boltzmann measure, and the stationary probability of observing a configuration $\phi(x)$ is given by $P_s[\phi]\propto \exp(-\mathcal F[\phi]/\epsilon)$. In this case, the relative stability of the phases $\phi_H$ and $\phi_I$ can be inferred from the values of $\mathcal F[\phi_H]$ and $\mathcal F[\phi_I]$. Transitions between these states involve a reaction path passing through a saddle-point configuration on $\mathcal F[\phi]$.

In contrast, when $\lambda\not=0$, TRS is broken because 
$\mu$ does not satisfy the Schwarz condition on its functional derivative 
~\cite{grafke_cates2017,obyrne2020,obyrne2023}. Consequently, the stationary distribution of the system is no longer available. Therefore, $\mathcal F[\phi]$ provides no information on the relative stability of $\phi_H$ and $\phi_I$. Instead, characterizing their relative stability relies on dynamics.

\textit{Phase transitions and quasipotential--} 
We use Freidlin-Wentzell large-deviation theory (LDT) to compute transition rates from $\phi_H$ to $\phi_I$ and vice versa, along with most likely paths~\cite{freidlinWentzell1998}, in the limit as $\epsilon\to 0$ (when the system is either in $\phi_H$ or $\phi_I$ with probability one, and proper phases can be defined). Denoting $k_{I,H}$ as the rate to transition from $\phi_I$ to $\phi_H$, it is given by $k_{I,H}\asymp \exp\left( -V_{\phi_I}(\phi_H)/\epsilon\right)$, where $V_{\phi_I}(\phi_H)$ is the quasipotential of $\phi_H$ relative to $\phi_I$, akin to a potential barrier in Arrhenius’ law. A similar expression holds for $k_{H,I}$, the rate from $\phi_H$ to $\phi_I$. Assessing the relative stability of the phases relies on the difference in the logarithm of the escape rates:
\begin{align}
\epsilon \log k_{I,H}- \epsilon \log k_{H,I} \asymp -V_{\phi_I}(\phi_H)+V_{\phi_H}(\phi_I),
\label{eq:escape_rates_ratio}
\end{align} 
This expression is positive when $\phi_H$ is preferred and negative when $\phi_I$ is.
The quasipotential values $V_{\phi_I}(\phi_H)$ and $V_{\phi_H}(\phi_I)$ depend on system control parameters like $\lambda$ and $\phi_0$, thus their difference can switch sign, indicating a first-order phase transition (FOPT). This allows for analyzing these transitions by computing these quasipotentials for various $\lambda$ and $\phi_0$ values, as suggested in~\cite{zakine2022}. These quasipotentials are obtained as minima of the action functional $S_T[\phi]$, defined as:
\begin{equation}
    \label{eq:action}
    S_T[\phi] = \int_0^T \int_\Omega |\nabla^{-1}( \partial_t\phi -\nabla^2 \mu)|^2dx dt
\end{equation}
where $\Omega$ denotes the domain. Minimizing action~\eqref{eq:action} with respect to both $T$ and $\phi$, subject to 
$\phi(t=0,x) = \phi_H$ and $\phi(t=T,x) = \phi_I$ yields $V_{\phi_H}(\phi_I)$, and subject to $\phi(t=0,x) = \phi_I$ and $\phi(t=T,x) = \phi_H$
 yields $V_{\phi_I}(\phi_H)$.

\textit{Deep gMAM--} 
The key feature of the method, introduced in \cite{simonnet_computing_2023}, is to replace the field $\phi(x,t)$ 
with an ansatz satisfying
the spatio-temporal boundary conditions and involving deep neural
networks. The minimization of \eqref{eq:action} is achieved
by a stochastic gradient descent (SGD) algorithm where space-time
collocation points are randomly drawn at each SGD step.
Such procedure is very often used in problems involving PINNs.
The method is simple to implement, highly flexible, and, importantly, able to tackle problems in higher dimensions not accessible by classical approaches. Furthermore, it provides an analytical parametric approximation of the various fields across the entire spatio-temporal domain.

In this study, results from the deep gMAM algorithm in $d=1$ were validated against those from classical gMAM implementation, which discretizes the field in space and time and is somewhat more intricate. For further details on both algorithms, especially regarding optimization on $T$ via reparametrization of the solution using arc-length $s$ instead of physical time $t$, we refer to the Supplemental Material (SM)~\footnote{The Supplemental Material can be found at [INSERT LINK] and includes additional references~\cite{donnelly_symplectic_2005,forest_fourth_order_1990,ERenVE2002, Cai2, adam}. The codes can be found at \url{https://github.com/rzakine/gMAM_active-Model-B}.}.

\textit{Phase diagram in 1d--} We focus first on the one-dimensional system, whose dynamics reads
\begin{align}
\partial_t\phi = -\partial_x^2[ \partial_x^2\phi +\phi-\phi^3-\lambda(\partial_x\phi)^2] + \partial_x\xi, 
\label{eq:SPDE_amb}
\end{align}
with $\langle \xi(x,t)\xi(x',t')\rangle =2\epsilon \delta(t-t')\delta(x-x')$. Space has been rescaled such that all lengths are given in units of $\sqrt{\nu}$. We consider a system of size $L$ and we take periodic boundary conditions. The relevant parameters are thus $L$, the total mass $\phi_0\equiv L^{-1}\int_0^L \phi~dx$, and the activity level $\lambda$.
The constant density solution of Eq.~\eqref{eq:SPDE_amb} is the homogeneous state $\phi_H$, and since the mass
$\phi_0$ is conserved, we have $\phi_H=\phi_0$.
We restrain the study to the region $\phi_0>0$, since Eq.~\eqref{eq:SPDE_amb} is invariant under $(\lambda,\phi)\to(-\lambda,-\phi)$. The homogeneous state $\phi_H$ is always a stable fixed point of the noiseless dynamics for $\phi_0>\phi_\mathrm{sp^+}^\lambda$, where $\phi_\mathrm{sp^+}^\lambda=1/\sqrt{3}$ is the frontier of the spinodal in the space $(\lambda,\phi_0)$ for $\phi_0>0$. We are interested in the region where $\phi_H$ competes with the inhomogeneous state $\phi_I$. In the infinite system size limit, this region lies between the spinodal $\phi_\mathrm{sp^+}^\lambda$ (red line in Fig.~\ref{fig:phase_diagram_amB}(b)) and the binodal curve $\phi_\mathrm{bi^+}^\lambda$ (black line in Fig.~\ref{fig:phase_diagram_amB}(b-c)) that yields the bulk densities of each phase when the system undergoes a phase separation~\cite{wittkowski2014,solon2018binodal}. 
We will denote by $\phi_\mathrm{f.o.}^\lambda$ the transition density indicating the change of thermodynamic stability of the two competing metastable states, $\phi_I$ and $\phi_H$. Naturally we have $\phi_\mathrm{sp^+}^\lambda\leq\phi_\mathrm{f.o.}^\lambda\leq\phi_\mathrm{bi^+}^\lambda$.

First, let us recall that for large but finite systems, the phase-separated state  cannot be the preferred phase if $\phi_0$ is taken too close to the binodal density $\phi_\mathrm{bi^+}^\lambda$. For instance, in equilibrium, (i.e. $\lambda=0$) the binodal densities are $\phi_\mathrm{bi^\pm}^{\lambda=0}=\pm 1$ but a free energy argument that compares interfaces and bulk contributions shows that $\phi_\mathrm{f.o.}^{\lambda=0}$ converges to 1 as $\phi_\mathrm{f.o.}^{\lambda=0}\sim 1-(1/L)^{1/2}$. More than that, due to finite-size effects, $\phi_I$ may not exist at all when there is not enough space in the domain to nucleate the phase separation. Hence, one should keep in mind that in a finite system, say of size $L$, bistability can only be observed below some threshold density $\phi_{\mathrm{m}_L^+}^{\lambda=0}\leq \phi_\mathrm{bi^+}^{\lambda=0}$, represented as the blue curve in Fig.~\ref{fig:phase_diagram_amB}. 
Nonetheless, we have $\phi_{\mathrm{m}_L^\pm}^\lambda\to\phi_\mathrm{bi^\pm}^\lambda$ as $L\to \infty$. To pinpoint the FOPT, we run the gMAM algorithm for $\phi_0\in[\phi_\mathrm{sp}^{\lambda},\phi_{\mathrm{m}_L^+}^{\lambda}]$ and $\lambda\in[-10,10]$. Solving $V_{\phi_H}(\phi_I)=V_{\phi_I}(\phi_H)$ identifies the FOPT line $\phi_\mathrm{f.o.}^\lambda$, the purple dashed line in Fig.~\ref{fig:phase_diagram_amB}(c), which splits the diagram into two regions: for $\phi_0<\phi_\mathrm{f.o.}^\lambda$ the thermodynamically stable state is the inhomogeneous one, $\phi_I$, while for $\phi_0>\phi_\mathrm{f.o.}^\lambda$ the homogeneous state $\phi_H=\phi_0$ is preferred. Interestingly, we also find that the binodal and the FOPT have a reentrance direction along $\lambda$ that does not exist in the system of infinite size (see Fig.~\ref{fig:phase_diagram_amB}(c)).

\begin{figure}
\includegraphics[width=\columnwidth]{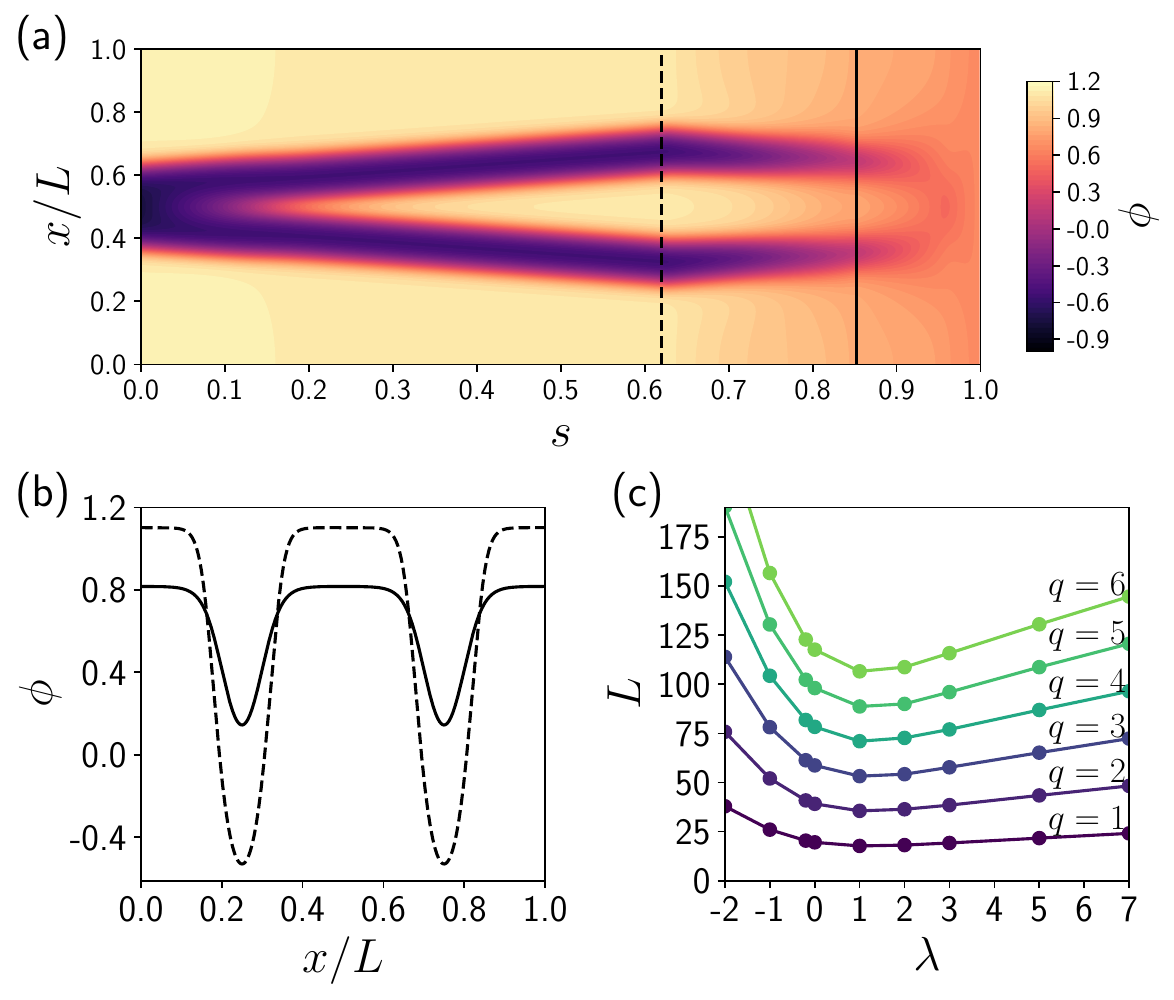}
  \caption{
   (a) Minimum action path joining $\phi_I$ (at $s=0$) to $\phi_H$ (at $s=1$) for $\lambda=2$, $\phi_0=0.65$ and $ L=44.7$ in $d=1$ dimension. The vertical lines pinpoint the states where the norm of the flow is minimal (and almost zero), corresponding to the states close to the critical points. The corresponding critical points are displayed in panel (b).
   The state at the dashed line lies in the basin of attraction of the inhomogeneous state, while the state at the solid line lies on the separatrix between the $\phi_I$ and $\phi_H$. The action from the dashed line to the solid line is strictly positive, while the action from the solid line to $\phi_H$ is zero.
   (b) Pair of critical states displaying two bumps, for same parameters as panel (a).  If $L=L_2^\star$, these two states merge in a saddle-node bifurcation.
  (c) Threshold lengths $L^\star_q(\lambda)$ indicating the apparition of critical states with a given number $q$ of bumps as a function of  the system activity $\lambda$. Above the critical $q$-line, pairs of critical states with $q$ bumps are dynamically accessible.
  }
  \label{fig:path_and_trend_criticalPoints}
\end{figure}

\textit{Reaction paths in 1d--}
We consider first the reaction path starting from the homogeneous state $\phi_H$ and reaching $\phi_I$, and we compute $V_{\phi_H}(\phi_I)$ for different values of $\lambda$ and system size $L$. Interestingly  this path is very close to the heteroclinic orbit joining $\phi_H$ to $\phi_I$, and going through the critical (saddle) state $\phi_{c,1}(x)$ that displays one density bump (see Fig.~\ref{fig:phase_diagram_amB}(a)) and  possesses only one unstable direction. This behavior is very similar to the equilibrium nucleation scenario occurring in the Cahn-Hilliard dynamics, as already noted in~\cite{cates_classical_2022}: to escape $\phi_H$, the system only needs to nucleate a finite size droplet of the opposite phase. The cost for the action associated to this event is always finite, and the value of the action does not differ much from the one computed using the time-reversed relaxational path (a few percent difference, not shown).

\begin{figure}
  \includegraphics[width=1\columnwidth]{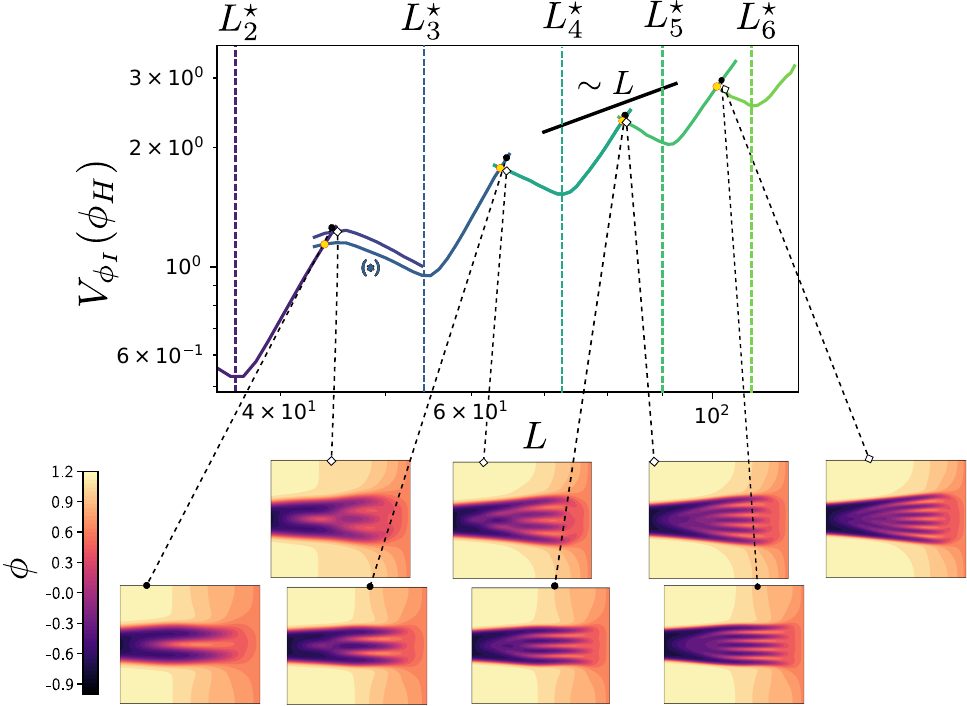}
  \caption{Minimum action $V_{\phi_I}(\phi_H)$ as a function of the system size $L$ (top panel), for paths starting at $\phi_I$ and reaching $\phi_H$. Here $\lambda=2$ and $\phi_0=0.65$. The action non-monotonically increases because increasing the system size $L$ allows for qualitatively different reaction paths. The successive branches of the curve correspond to different types of paths displaying an increasing number of bumps, see bottom panels. 
  The yellow dots indicate where branches cross each other. The $(*)$ symbol indicates a branch on which the path is no longer axisymmetric (see SM). The vertical dashed lines indicate the critical lengths $L_q^\star$, also given in Fig.~\ref{fig:path_and_trend_criticalPoints}(c).
  }
  \label{fig:scaling_AMB}
\end{figure}

In contrast, the transition from $\phi_I$ to $\phi_H$ is more complex, and its analysis had never been explored so far. For $\phi_0>0$, as $\lambda$ increases, the reaction path no longer follows the time-reverse relaxation path that goes through the saddle $\phi_{c,1}$, but rather passes close to critical points with a large number of unstable directions, see Fig.~\ref{fig:path_and_trend_criticalPoints}(a) and \ref{fig:path_and_trend_criticalPoints}(b), as it may sometimes be observed in nonequilibrium systems~\cite{zakine2022,simonnet_computing_2023}. 
Any critical points $\phi_c$ can be obtained by solving the noiseless and stationary version of Eq.~\eqref{eq:SPDE_amb}. It solves $\partial_x^2\phi_c=-\phi_c+\phi_c^3+\lambda(\partial_x\phi_c)^2 + \mu_0$, with $\mu_0$ a constant, $L^{-1}\int_0^L \phi_c(x)dx=\phi_0$, and $\phi_c$ subject to the constraints of periodicity. A Newton mapping similar to the one introduced in \cite{wittkowski2014} enables us to compute precisely the critical points using a symplectic scheme (see SM). For given $\lambda$ and $\phi_0$, pairs of critical points with $q$ bumps ($q\in \mathbb N^*$) appear at critical values of the system size denoted $L_q^\star$, reported in Fig.~\ref{fig:path_and_trend_criticalPoints}(c). The saddle-node bifurcation at $L_q^\star$ occurs when the system size $L$ is large enough to fit an additional bump on the density profile. For $L=L_{q}^\star$, one degenerate critical state $\phi^\star_{c,q}$ becomes accessible to the dynamics. As $L>L_q^\star$, the degeneracy is lifted and two distinct critical states of $q$ bumps appear. Any of the states $\phi_{c,q}$ can be decomposed into $q$ identical bumps of size $L/q$. 
In particular, the state with bumps of largest amplitude strictly lies in the basin of attraction of $\phi_I$, while the other state lies on the separatrix between $\phi_I$ and $\phi_H$. We display an example of such a pair of critical states for $q=2$ in Fig.~\ref{fig:path_and_trend_criticalPoints}(b). For all $q\geq 2$, the critical states are of Morse index $q\geq 2$. The case $q=1$ is special as it corresponds to the apparition of the inhomogeneous metastable state $\phi_I$, jointly with the critical state of Morse index 1, $\phi_{c,1}(x)$. A sketch of the structure of the deterministic flow between critical points is given in the SM.
In summary, while the path from $\phi_H$ to $\phi_I$ indeed resembles the equilibrium one, the path from $\phi_I$ to $\phi_H$ displays spatial microstructures which are not present in equilibrium. Notably, the number $q$ of bumps along the instanton changes with $L$, see Fig.~\ref{fig:scaling_AMB}, but also depends on $\phi_0$ and $\lambda$, as indicated by the $\mathcal I_q$-labeled regions in Fig.~\ref{fig:phase_diagram_amB}.
In the SM, we provide a more detailed discussion on the paths selection, and we show that the number of bumps along the path cannot be simply obtained from a spectral analysis.

\begin{figure}
\centering
\includegraphics[width=0.99\columnwidth]{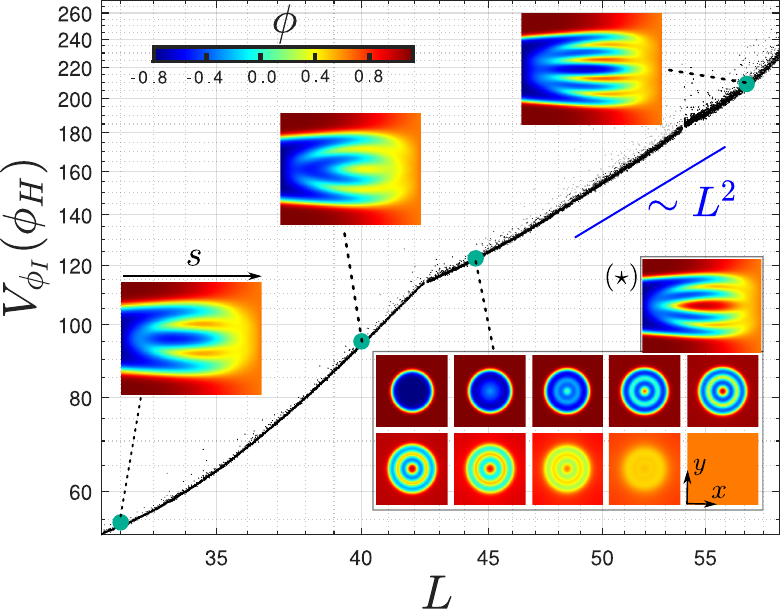}
\caption{Minimum action $V_{\phi_I}(\phi_H)$ as a function of $L$
obtained using the deep gMAM method, in $d=2$ dimensions. The bottom-right panels shows the successive states along the Minimum Action Path joining $\phi_I$ to $\phi_H$ for $L = 44.7, \lambda = 2, \phi_0 = 0.65$. The $(\star)$ panel is the same solution in $(s,r)$ coordinates where $r$ is the radial coordinate and $s$ is the arclength coordinate. Other reaction paths in radial coordinates are shown for different values of $L = 32, 40, 57$. They exhibit additional microstructures as $L$ increases.
}
\label{fig:instanton_2d}
\end{figure}

\begin{figure}
\centering
\includegraphics[width=1\columnwidth]{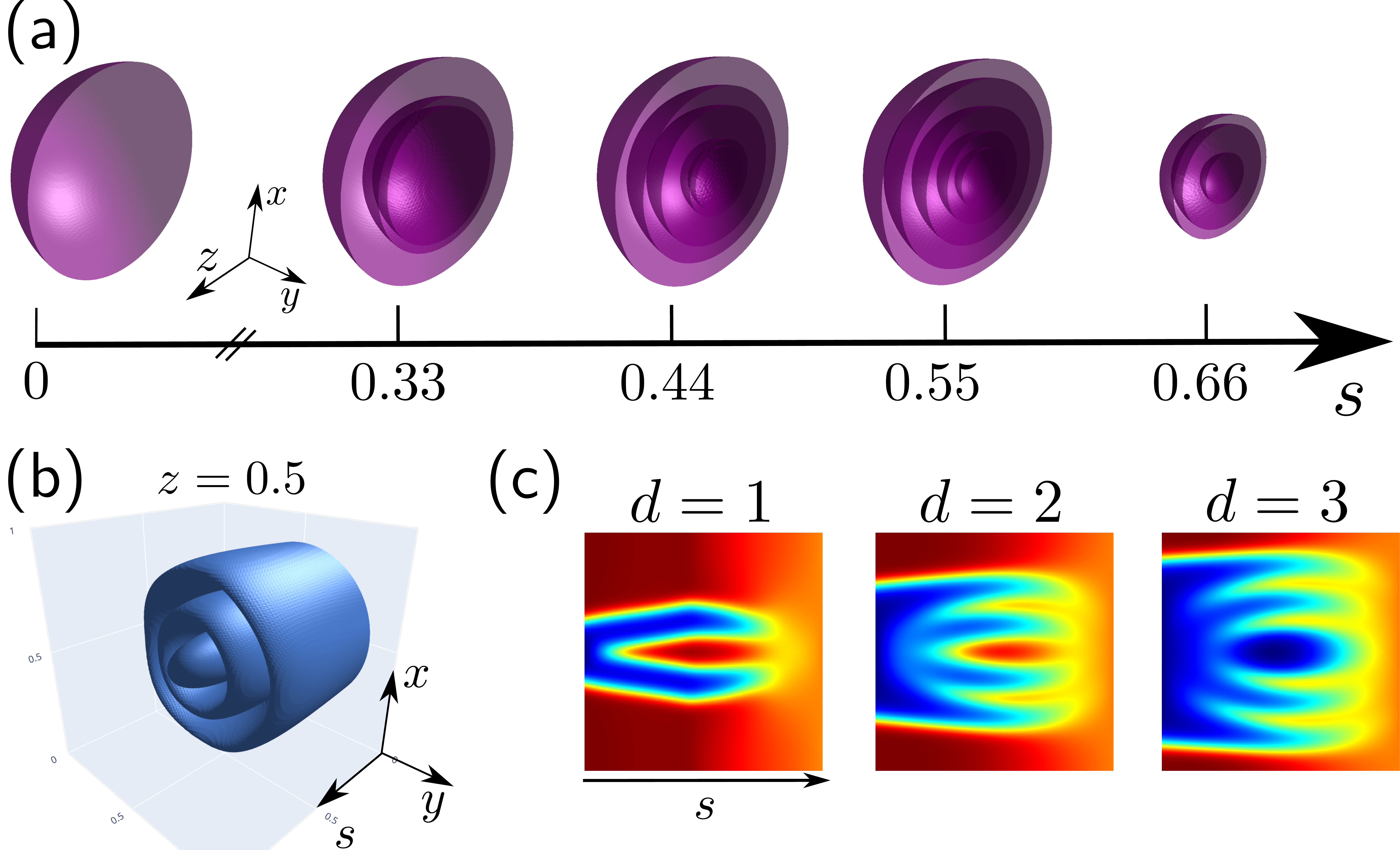}
\caption{(a) Isosurfaces $\phi(x,y,z)=0$ along the reaction path from $\phi_I$ to $\phi_H$, for $d=3$. A slice is shown at $y=0.5L$. (b) Isosurface $\phi=0$ in space $(s,x,y)$ and $z=0.5L$. (c) Comparison of the reaction paths in space $(s,r)$ for $L=44.7$ for $d=1$, $2$, and $3$. For all panels: $\lambda=2$, $\phi_0=0.65$.
}
\label{fig:3d_plots}
\end{figure}

\textit{Phase transitions in 2d and 3d--} 
The reaction paths are also computable in higher dimensions using the deep gMAM algorithm. We specifically examine transitions from $\phi_I$ to $\phi_H$, having also checked that transitions from $\phi_H$ to $\phi_I$ are in line with classic nucleation theory~\cite{cates_classical_2022} (results not shown). In $d=2$, we investigate the dependence on domain size $L$, as illustrated in Fig.~\ref{fig:instanton_2d}. These transitions display radial symmetry, with microstructures increasing as $L$ grows. The system exhibits extensive action values $V_{\phi_I}(\phi_H)$, scaling as $\sim L^2$. Although the action in Fig.~\ref{fig:instanton_2d} seems monotonic, rescaling it by $L^2$ reveals a very similar non-monotonic behavior as in one dimension (not shown).
Evidence suggests that instantons do not traverse multi-spike profiles, which are numerically identified as critical states of the AMB (see SM and Ref.~\cite{bates2000}), as their action values consistently exceed those of the radially symmetric path. Additionally, we compute 3D transitions, with one typical example shown in Fig.~\ref{fig:3d_plots}. These transitions also exhibit radial symmetry. Notably, there is a significant dimensional effect: for same extension $L$, microstructures have more room to span as dimension increases.
This is anticipated due to mass conservation, as the positive mass concentrate towards the domain's corners. 
It is noteworthy that in dimensions $d\geq 2$, characterizing critical states in Cahn-Hilliard is more challenging~\cite{bates2000} than in $d=1$~\cite{bates1993}, and this question remains open for the AMB. Overall, comparing to the Arrhenius law for $\lambda=0$ reveals that the active term substantially reduces the action required to escape the inhomogeneous state.

\textit{Conclusion--} 
We have computed the phase diagram of the AMB in $d=1$, identified various nucleation scenarios in the binodal, and demonstrated similar instanton phenomenology in $d=2$ and $3$. By computing the reaction paths, we identified regions where the homogeneous state is thermodynamically preferred. The fact that the action $V_{\phi_I}(\phi_H)$
remains extensive in system size, while $V_{\phi_H}(\phi_I)$
remains finite, confirms that eventually, the system should phase-separate as $L\to\infty$ when lying in the binodal region.
Our results align with those of Cates and Nardini~\cite{cates_classical_2022}, who showed that nucleation from the homogeneous state in AMB for $d\geq 2$ is qualitatively similar to classical nucleation theory in equilibrium. Moreover, a common feature among all cases is the presence of microstructures with more complex patterns as the dimension $d$ increases.
These patterns help decreasing significantly the action required to escape the inhomogeneous state.
Our numerical results were obtained using the deep gMAM approach~\cite{simonnet_computing_2023} and cross-checked in $d=1$ by running the classical gMAM~\cite{VE2008gMAM}. While the latter algorithm is more accurate, the discretization scheme adopted for the Cahn-Hilliard equation becomes numerically prohibitive when $d \geq 2$. The deep gMAM approach does not suffer much from an increase in the dimension $d$. Consequently, we were able to compute transitions for $d=3$. Overall, these features make the proposed method relevant for numerous active matter systems that may undergo phase separation, and for field theories displaying metastable states in high dimension.

\vspace{5pt}
\begin{acknowledgments}
  {We thank  Thibaut Arnoulx de Pirey,  Cesare Nardini, Jeremy O'Byrne, and Julien Tailleur for interesting discussions. This work was supported by the Materials Research Science and Engineering Center (MRSEC) program of the National Science Foundation under Grants No. NSF DMR-1420073 and by Grant No. NSF DMR-1710163.
  R. Z. thanks Laboratoire MSC Paris for hospitality. R. Z. and E. V.-E. would also like to thank the Center for Data Science ENS Paris for hospitality.}
\end{acknowledgments}

\bibliographystyle{apsrev4-2}
\bibliography{phaseTransition_AMB}

\begin{thebibliography}{46}%
\makeatletter
\providecommand \@ifxundefined [1]{%
 \@ifx{#1\undefined}
}%
\providecommand \@ifnum [1]{%
 \ifnum #1\expandafter \@firstoftwo
 \else \expandafter \@secondoftwo
 \fi
}%
\providecommand \@ifx [1]{%
 \ifx #1\expandafter \@firstoftwo
 \else \expandafter \@secondoftwo
 \fi
}%
\providecommand \natexlab [1]{#1}%
\providecommand \enquote  [1]{``#1''}%
\providecommand \bibnamefont  [1]{#1}%
\providecommand \bibfnamefont [1]{#1}%
\providecommand \citenamefont [1]{#1}%
\providecommand \href@noop [0]{\@secondoftwo}%
\providecommand \href [0]{\begingroup \@sanitize@url \@href}%
\providecommand \@href[1]{\@@startlink{#1}\@@href}%
\providecommand \@@href[1]{\endgroup#1\@@endlink}%
\providecommand \@sanitize@url [0]{\catcode `\\12\catcode `\$12\catcode
  `\&12\catcode `\#12\catcode `\^12\catcode `\_12\catcode `\%12\relax}%
\providecommand \@@startlink[1]{}%
\providecommand \@@endlink[0]{}%
\providecommand \url  [0]{\begingroup\@sanitize@url \@url }%
\providecommand \@url [1]{\endgroup\@href {#1}{\urlprefix }}%
\providecommand \urlprefix  [0]{URL }%
\providecommand \Eprint [0]{\href }%
\providecommand \doibase [0]{https://doi.org/}%
\providecommand \selectlanguage [0]{\@gobble}%
\providecommand \bibinfo  [0]{\@secondoftwo}%
\providecommand \bibfield  [0]{\@secondoftwo}%
\providecommand \translation [1]{[#1]}%
\providecommand \BibitemOpen [0]{}%
\providecommand \bibitemStop [0]{}%
\providecommand \bibitemNoStop [0]{.\EOS\space}%
\providecommand \EOS [0]{\spacefactor3000\relax}%
\providecommand \BibitemShut  [1]{\csname bibitem#1\endcsname}%
\let\auto@bib@innerbib\@empty
\bibitem [{\citenamefont {Cates}\ and\ \citenamefont
  {Nardini}(2023)}]{cates_classical_2022}%
  \BibitemOpen
  \bibfield  {author} {\bibinfo {author} {\bibfnamefont {M.~E.}\ \bibnamefont
  {Cates}}\ and\ \bibinfo {author} {\bibfnamefont {C.}~\bibnamefont
  {Nardini}},\ }\href {https://doi.org/10.1103/PhysRevLett.130.098203}
  {\bibfield  {journal} {\bibinfo  {journal} {Phys. Rev. Lett.}\ }\textbf
  {\bibinfo {volume} {130}},\ \bibinfo {pages} {098203} (\bibinfo {year}
  {2023})}\BibitemShut {NoStop}%
\bibitem [{\citenamefont {Torrie}\ and\ \citenamefont
  {Valleau}(1977)}]{torrie_nonphysical_1977}%
  \BibitemOpen
  \bibfield  {author} {\bibinfo {author} {\bibfnamefont {G.~M.}\ \bibnamefont
  {Torrie}}\ and\ \bibinfo {author} {\bibfnamefont {J.~P.}\ \bibnamefont
  {Valleau}},\ }\href {https://doi.org/10.1016/0021-9991(77)90121-8} {\bibfield
   {journal} {\bibinfo  {journal} {Journal of Computational Physics}\ }\textbf
  {\bibinfo {volume} {23}},\ \bibinfo {pages} {187} (\bibinfo {year}
  {1977})}\BibitemShut {NoStop}%
\bibitem [{\citenamefont {Bolhuis}\ \emph {et~al.}(2002)\citenamefont
  {Bolhuis}, \citenamefont {Chandler}, \citenamefont {Dellago},\ and\
  \citenamefont {Geissler}}]{bolhuis2002}%
  \BibitemOpen
  \bibfield  {author} {\bibinfo {author} {\bibfnamefont {P.~G.}\ \bibnamefont
  {Bolhuis}}, \bibinfo {author} {\bibfnamefont {D.}~\bibnamefont {Chandler}},
  \bibinfo {author} {\bibfnamefont {C.}~\bibnamefont {Dellago}},\ and\ \bibinfo
  {author} {\bibfnamefont {P.~L.}\ \bibnamefont {Geissler}},\ }\href
  {https://doi.org/10.1146/annurev.physchem.53.082301.113146} {\bibfield
  {journal} {\bibinfo  {journal} {Annual Review of Physical Chemistry}\
  }\textbf {\bibinfo {volume} {53}},\ \bibinfo {pages} {291} (\bibinfo {year}
  {2002})}\BibitemShut {NoStop}%
\bibitem [{\citenamefont {Grafke}\ and\ \citenamefont
  {Vanden-Eijnden}(2019)}]{grafke_numerical_2019}%
  \BibitemOpen
  \bibfield  {author} {\bibinfo {author} {\bibfnamefont {T.}~\bibnamefont
  {Grafke}}\ and\ \bibinfo {author} {\bibfnamefont {E.}~\bibnamefont
  {Vanden-Eijnden}},\ }\href {https://doi.org/10.1063/1.5084025} {\bibfield
  {journal} {\bibinfo  {journal} {Chaos: An Interdisciplinary Journal of
  Nonlinear Science}\ }\textbf {\bibinfo {volume} {29}},\ \bibinfo {pages}
  {063118} (\bibinfo {year} {2019})}\BibitemShut {NoStop}%
\bibitem [{\citenamefont {Rotskoff}\ \emph {et~al.}(2022)\citenamefont
  {Rotskoff}, \citenamefont {Mitchell},\ and\ \citenamefont
  {Vanden-Eijnden}}]{rotskoff_active_2022}%
  \BibitemOpen
  \bibfield  {author} {\bibinfo {author} {\bibfnamefont {G.~M.}\ \bibnamefont
  {Rotskoff}}, \bibinfo {author} {\bibfnamefont {A.~R.}\ \bibnamefont
  {Mitchell}},\ and\ \bibinfo {author} {\bibfnamefont {E.}~\bibnamefont
  {Vanden-Eijnden}},\ }in\ \href
  {https://proceedings.mlr.press/v145/rotskoff22a.html} {\emph {\bibinfo
  {booktitle} {Proceedings of the 2nd {Mathematical} and {Scientific} {Machine}
  {Learning} {Conference}}}}\ (\bibinfo  {publisher} {PMLR},\ \bibinfo {year}
  {2022})\ pp.\ \bibinfo {pages} {757--780}\BibitemShut {NoStop}%
\bibitem [{\citenamefont {Binder}(1987)}]{binder1987}%
  \BibitemOpen
  \bibfield  {author} {\bibinfo {author} {\bibfnamefont {K.}~\bibnamefont
  {Binder}},\ }\href {https://doi.org/10.1088/0034-4885/50/7/001} {\bibfield
  {journal} {\bibinfo  {journal} {Reports on Progress in Physics}\ }\textbf
  {\bibinfo {volume} {50}},\ \bibinfo {pages} {783} (\bibinfo {year}
  {1987})}\BibitemShut {NoStop}%
\bibitem [{\citenamefont {Richard}\ \emph {et~al.}(2016)\citenamefont
  {Richard}, \citenamefont {Löwen},\ and\ \citenamefont
  {Speck}}]{richard_speck2016}%
  \BibitemOpen
  \bibfield  {author} {\bibinfo {author} {\bibfnamefont {D.}~\bibnamefont
  {Richard}}, \bibinfo {author} {\bibfnamefont {H.}~\bibnamefont {Löwen}},\
  and\ \bibinfo {author} {\bibfnamefont {T.}~\bibnamefont {Speck}},\ }\href
  {https://doi.org/10.1039/C6SM00485G} {\bibfield  {journal} {\bibinfo
  {journal} {Soft Matter}\ }\textbf {\bibinfo {volume} {12}},\ \bibinfo {pages}
  {5257} (\bibinfo {year} {2016})}\BibitemShut {NoStop}%
\bibitem [{\citenamefont {Omar}\ \emph {et~al.}(2021)\citenamefont {Omar},
  \citenamefont {Klymko}, \citenamefont {GrandPre},\ and\ \citenamefont
  {Geissler}}]{omar_phase_2021}%
  \BibitemOpen
  \bibfield  {author} {\bibinfo {author} {\bibfnamefont {A.~K.}\ \bibnamefont
  {Omar}}, \bibinfo {author} {\bibfnamefont {K.}~\bibnamefont {Klymko}},
  \bibinfo {author} {\bibfnamefont {T.}~\bibnamefont {GrandPre}},\ and\
  \bibinfo {author} {\bibfnamefont {P.~L.}\ \bibnamefont {Geissler}},\ }\href
  {https://doi.org/10.1103/PhysRevLett.126.188002} {\bibfield  {journal}
  {\bibinfo  {journal} {Physical Review Letters}\ }\textbf {\bibinfo {volume}
  {126}},\ \bibinfo {pages} {188002} (\bibinfo {year} {2021})}\BibitemShut
  {NoStop}%
\bibitem [{\citenamefont {Kramers}(1940)}]{kramers_brownian_1940}%
  \BibitemOpen
  \bibfield  {author} {\bibinfo {author} {\bibfnamefont {H.~A.}\ \bibnamefont
  {Kramers}},\ }\href {https://doi.org/10.1016/S0031-8914(40)90098-2}
  {\bibfield  {journal} {\bibinfo  {journal} {Physica}\ }\textbf {\bibinfo
  {volume} {7}},\ \bibinfo {pages} {284} (\bibinfo {year} {1940})}\BibitemShut
  {NoStop}%
\bibitem [{\citenamefont {Onsager}\ and\ \citenamefont
  {Machlup}(1953)}]{onsager1953}%
  \BibitemOpen
  \bibfield  {author} {\bibinfo {author} {\bibfnamefont {L.}~\bibnamefont
  {Onsager}}\ and\ \bibinfo {author} {\bibfnamefont {S.}~\bibnamefont
  {Machlup}},\ }\href {https://doi.org/10.1103/PhysRev.91.1505} {\bibfield
  {journal} {\bibinfo  {journal} {Physical Review}\ }\textbf {\bibinfo {volume}
  {91}},\ \bibinfo {pages} {1505} (\bibinfo {year} {1953})}\BibitemShut
  {NoStop}%
\bibitem [{\citenamefont {E}\ \emph {et~al.}(2007)\citenamefont {E},
  \citenamefont {Ren},\ and\ \citenamefont
  {Vanden-Eijnden}}]{e_simplified_2007}%
  \BibitemOpen
  \bibfield  {author} {\bibinfo {author} {\bibfnamefont {W.}~\bibnamefont {E}},
  \bibinfo {author} {\bibfnamefont {W.}~\bibnamefont {Ren}},\ and\ \bibinfo
  {author} {\bibfnamefont {E.}~\bibnamefont {Vanden-Eijnden}},\ }\href
  {https://doi.org/10.1063/1.2720838} {\bibfield  {journal} {\bibinfo
  {journal} {The Journal of Chemical Physics}\ }\textbf {\bibinfo {volume}
  {126}},\ \bibinfo {pages} {164103} (\bibinfo {year} {2007})}\BibitemShut
  {NoStop}%
\bibitem [{\citenamefont {Freidlin}\ and\ \citenamefont
  {Wentzell}(1998)}]{freidlinWentzell1998}%
  \BibitemOpen
  \bibfield  {author} {\bibinfo {author} {\bibfnamefont {M.~I.}\ \bibnamefont
  {Freidlin}}\ and\ \bibinfo {author} {\bibfnamefont {A.~D.}\ \bibnamefont
  {Wentzell}},\ }\href {https://doi.org/10.1007/978-1-4612-0611-8} {\emph
  {\bibinfo {title} {Random {Perturbations} of {Dynamical} {Systems}}}},\
  \bibinfo {series} {Grundlehren der mathematischen {Wissenschaften}}, Vol.\
  \bibinfo {volume} {260}\ (\bibinfo  {publisher} {Springer New York},\
  \bibinfo {address} {New York, NY},\ \bibinfo {year} {1998})\BibitemShut
  {NoStop}%
\bibitem [{\citenamefont {Ludwig}(1975)}]{ludwig1975}%
  \BibitemOpen
  \bibfield  {author} {\bibinfo {author} {\bibfnamefont {D.}~\bibnamefont
  {Ludwig}},\ }\href {https://doi.org/10.1137/1017070} {\bibfield  {journal}
  {\bibinfo  {journal} {SIAM Review}\ }\textbf {\bibinfo {volume} {17}},\
  \bibinfo {pages} {605} (\bibinfo {year} {1975})}\BibitemShut {NoStop}%
\bibitem [{\citenamefont {Maier}\ and\ \citenamefont
  {Stein}(1992)}]{maier_stein1992}%
  \BibitemOpen
  \bibfield  {author} {\bibinfo {author} {\bibfnamefont {R.~S.}\ \bibnamefont
  {Maier}}\ and\ \bibinfo {author} {\bibfnamefont {D.~L.}\ \bibnamefont
  {Stein}},\ }\href {https://doi.org/10.1103/PhysRevLett.69.3691} {\bibfield
  {journal} {\bibinfo  {journal} {Phys. Rev. Lett.}\ }\textbf {\bibinfo
  {volume} {69}},\ \bibinfo {pages} {3691} (\bibinfo {year}
  {1992})}\BibitemShut {NoStop}%
\bibitem [{\citenamefont {Dykman}\ \emph {et~al.}(1994)\citenamefont {Dykman},
  \citenamefont {Mori}, \citenamefont {Ross},\ and\ \citenamefont
  {Hunt}}]{dykman_large_1994}%
  \BibitemOpen
  \bibfield  {author} {\bibinfo {author} {\bibfnamefont {M.~I.}\ \bibnamefont
  {Dykman}}, \bibinfo {author} {\bibfnamefont {E.}~\bibnamefont {Mori}},
  \bibinfo {author} {\bibfnamefont {J.}~\bibnamefont {Ross}},\ and\ \bibinfo
  {author} {\bibfnamefont {P.~M.}\ \bibnamefont {Hunt}},\ }\href
  {https://doi.org/10.1063/1.467139} {\bibfield  {journal} {\bibinfo  {journal}
  {The Journal of Chemical Physics}\ }\textbf {\bibinfo {volume} {100}},\
  \bibinfo {pages} {5735} (\bibinfo {year} {1994})}\BibitemShut {NoStop}%
\bibitem [{\citenamefont {Bertini}\ \emph {et~al.}(2015)\citenamefont
  {Bertini}, \citenamefont {De~Sole}, \citenamefont {Gabrielli}, \citenamefont
  {Jona-Lasinio},\ and\ \citenamefont {Landim}}]{bertini2015}%
  \BibitemOpen
  \bibfield  {author} {\bibinfo {author} {\bibfnamefont {L.}~\bibnamefont
  {Bertini}}, \bibinfo {author} {\bibfnamefont {A.}~\bibnamefont {De~Sole}},
  \bibinfo {author} {\bibfnamefont {D.}~\bibnamefont {Gabrielli}}, \bibinfo
  {author} {\bibfnamefont {G.}~\bibnamefont {Jona-Lasinio}},\ and\ \bibinfo
  {author} {\bibfnamefont {C.}~\bibnamefont {Landim}},\ }\href
  {https://doi.org/10.1103/RevModPhys.87.593} {\bibfield  {journal} {\bibinfo
  {journal} {Reviews of Modern Physics}\ }\textbf {\bibinfo {volume} {87}},\
  \bibinfo {pages} {593} (\bibinfo {year} {2015})}\BibitemShut {NoStop}%
\bibitem [{\citenamefont {Bouchet}\ and\ \citenamefont
  {Reygner}(2016)}]{bouchet_generalisation_2016}%
  \BibitemOpen
  \bibfield  {author} {\bibinfo {author} {\bibfnamefont {F.}~\bibnamefont
  {Bouchet}}\ and\ \bibinfo {author} {\bibfnamefont {J.}~\bibnamefont
  {Reygner}},\ }\href {https://doi.org/10.1007/s00023-016-0507-4} {\bibfield
  {journal} {\bibinfo  {journal} {Annales Henri Poincaré}\ }\textbf {\bibinfo
  {volume} {17}},\ \bibinfo {pages} {3499} (\bibinfo {year}
  {2016})}\BibitemShut {NoStop}%
\bibitem [{\citenamefont {Grafke}\ \emph
  {et~al.}(2017{\natexlab{a}})\citenamefont {Grafke}, \citenamefont
  {Schäfer},\ and\ \citenamefont {Vanden-Eijnden}}]{grafke_long_2017}%
  \BibitemOpen
  \bibfield  {author} {\bibinfo {author} {\bibfnamefont {T.}~\bibnamefont
  {Grafke}}, \bibinfo {author} {\bibfnamefont {T.}~\bibnamefont {Schäfer}},\
  and\ \bibinfo {author} {\bibfnamefont {E.}~\bibnamefont {Vanden-Eijnden}},\
  }in\ \href {https://doi.org/10.1007/978-1-4939-6969-2_2} {\emph {\bibinfo
  {booktitle} {Recent {Progress} and {Modern} {Challenges} in {Applied}
  {Mathematics}, {Modeling} and {Computational} {Science}}}},\ \bibinfo
  {editor} {edited by\ \bibinfo {editor} {\bibfnamefont {R.}~\bibnamefont
  {Melnik}}, \bibinfo {editor} {\bibfnamefont {R.}~\bibnamefont {Makarov}},\
  and\ \bibinfo {editor} {\bibfnamefont {J.}~\bibnamefont {Belair}}}\ (\bibinfo
   {publisher} {Springer New York},\ \bibinfo {address} {New York, NY},\
  \bibinfo {year} {2017})\ pp.\ \bibinfo {pages} {17--55}\BibitemShut {NoStop}%
\bibitem [{\citenamefont {Woillez}\ \emph {et~al.}(2019)\citenamefont
  {Woillez}, \citenamefont {Zhao}, \citenamefont {Kafri}, \citenamefont
  {Lecomte},\ and\ \citenamefont {Tailleur}}]{woillez2019}%
  \BibitemOpen
  \bibfield  {author} {\bibinfo {author} {\bibfnamefont {E.}~\bibnamefont
  {Woillez}}, \bibinfo {author} {\bibfnamefont {Y.}~\bibnamefont {Zhao}},
  \bibinfo {author} {\bibfnamefont {Y.}~\bibnamefont {Kafri}}, \bibinfo
  {author} {\bibfnamefont {V.}~\bibnamefont {Lecomte}},\ and\ \bibinfo {author}
  {\bibfnamefont {J.}~\bibnamefont {Tailleur}},\ }\href
  {https://doi.org/10.1103/PhysRevLett.122.258001} {\bibfield  {journal}
  {\bibinfo  {journal} {Phys. Rev. Lett.}\ }\textbf {\bibinfo {volume} {122}},\
  \bibinfo {pages} {258001} (\bibinfo {year} {2019})}\BibitemShut {NoStop}%
\bibitem [{\citenamefont {Kardar}\ \emph {et~al.}(1986)\citenamefont {Kardar},
  \citenamefont {Parisi},\ and\ \citenamefont {Zhang}}]{kardar1986}%
  \BibitemOpen
  \bibfield  {author} {\bibinfo {author} {\bibfnamefont {M.}~\bibnamefont
  {Kardar}}, \bibinfo {author} {\bibfnamefont {G.}~\bibnamefont {Parisi}},\
  and\ \bibinfo {author} {\bibfnamefont {Y.-C.}\ \bibnamefont {Zhang}},\ }\href
  {https://doi.org/10.1103/PhysRevLett.56.889} {\bibfield  {journal} {\bibinfo
  {journal} {Physical Review Letters}\ }\textbf {\bibinfo {volume} {56}},\
  \bibinfo {pages} {889} (\bibinfo {year} {1986})}\BibitemShut {NoStop}%
\bibitem [{\citenamefont {Sun}\ \emph {et~al.}(1989)\citenamefont {Sun},
  \citenamefont {Guo},\ and\ \citenamefont {Grant}}]{sun_CKPZ_1989}%
  \BibitemOpen
  \bibfield  {author} {\bibinfo {author} {\bibfnamefont {T.}~\bibnamefont
  {Sun}}, \bibinfo {author} {\bibfnamefont {H.}~\bibnamefont {Guo}},\ and\
  \bibinfo {author} {\bibfnamefont {M.}~\bibnamefont {Grant}},\ }\href
  {https://doi.org/10.1103/PhysRevA.40.6763} {\bibfield  {journal} {\bibinfo
  {journal} {Physical Review A}\ }\textbf {\bibinfo {volume} {40}},\ \bibinfo
  {pages} {6763} (\bibinfo {year} {1989})}\BibitemShut {NoStop}%
\bibitem [{\citenamefont {Wittkowski}\ \emph {et~al.}(2014)\citenamefont
  {Wittkowski}, \citenamefont {Tiribocchi}, \citenamefont {Stenhammar},
  \citenamefont {Allen}, \citenamefont {Marenduzzo},\ and\ \citenamefont
  {Cates}}]{wittkowski2014}%
  \BibitemOpen
  \bibfield  {author} {\bibinfo {author} {\bibfnamefont {R.}~\bibnamefont
  {Wittkowski}}, \bibinfo {author} {\bibfnamefont {A.}~\bibnamefont
  {Tiribocchi}}, \bibinfo {author} {\bibfnamefont {J.}~\bibnamefont
  {Stenhammar}}, \bibinfo {author} {\bibfnamefont {R.~J.}\ \bibnamefont
  {Allen}}, \bibinfo {author} {\bibfnamefont {D.}~\bibnamefont {Marenduzzo}},\
  and\ \bibinfo {author} {\bibfnamefont {M.~E.}\ \bibnamefont {Cates}},\ }\href
  {https://doi.org/10.1038/ncomms5351} {\bibfield  {journal} {\bibinfo
  {journal} {Nature Communications}\ }\textbf {\bibinfo {volume} {5}},\
  \bibinfo {pages} {4351} (\bibinfo {year} {2014})}\BibitemShut {NoStop}%
\bibitem [{\citenamefont {Nardini}\ \emph {et~al.}(2017)\citenamefont
  {Nardini}, \citenamefont {Fodor}, \citenamefont {Tjhung}, \citenamefont {van
  Wijland}, \citenamefont {Tailleur},\ and\ \citenamefont
  {Cates}}]{nardini2017prx}%
  \BibitemOpen
  \bibfield  {author} {\bibinfo {author} {\bibfnamefont {C.}~\bibnamefont
  {Nardini}}, \bibinfo {author} {\bibfnamefont {E.}~\bibnamefont {Fodor}},
  \bibinfo {author} {\bibfnamefont {E.}~\bibnamefont {Tjhung}}, \bibinfo
  {author} {\bibfnamefont {F.}~\bibnamefont {van Wijland}}, \bibinfo {author}
  {\bibfnamefont {J.}~\bibnamefont {Tailleur}},\ and\ \bibinfo {author}
  {\bibfnamefont {M.~E.}\ \bibnamefont {Cates}},\ }\href
  {https://doi.org/10.1103/PhysRevX.7.021007} {\bibfield  {journal} {\bibinfo
  {journal} {Phys. Rev. X}\ }\textbf {\bibinfo {volume} {7}},\ \bibinfo {pages}
  {021007} (\bibinfo {year} {2017})}\BibitemShut {NoStop}%
\bibitem [{\citenamefont {Cates}(2019)}]{catesHouches2019}%
  \BibitemOpen
  \bibfield  {author} {\bibinfo {author} {\bibfnamefont {M.~E.}\ \bibnamefont
  {Cates}},\ }in\ \href@noop {} {\emph {\bibinfo {booktitle} {Lecture Notes of
  the Les Houches Summer School: Volume 112, September 2018}}},\ \bibinfo
  {editor} {edited by\ \bibinfo {editor} {\bibfnamefont {J.}~\bibnamefont
  {Tailleur}}, \bibinfo {editor} {\bibfnamefont {G.}~\bibnamefont {Gompper}},
  \bibinfo {editor} {\bibfnamefont {C.}~\bibnamefont {Marchetti}}, \bibinfo
  {editor} {\bibfnamefont {J.~M.}\ \bibnamefont {Yeomans}},\ and\ \bibinfo
  {editor} {\bibfnamefont {C.}~\bibnamefont {Salomon}}}\ (\bibinfo  {publisher}
  {Oxford University Press},\ \bibinfo {year} {2019})\BibitemShut {NoStop}%
\bibitem [{\citenamefont {O'Byrne}(2023)}]{obyrne2023}%
  \BibitemOpen
  \bibfield  {author} {\bibinfo {author} {\bibfnamefont {J.}~\bibnamefont
  {O'Byrne}},\ }\href {https://doi.org/10.1103/PhysRevE.107.054105} {\bibfield
  {journal} {\bibinfo  {journal} {Phys. Rev. E}\ }\textbf {\bibinfo {volume}
  {107}},\ \bibinfo {pages} {054105} (\bibinfo {year} {2023})}\BibitemShut
  {NoStop}%
\bibitem [{\citenamefont {Cates}\ and\ \citenamefont
  {Tailleur}(2015)}]{cates_motility_induced_2015}%
  \BibitemOpen
  \bibfield  {author} {\bibinfo {author} {\bibfnamefont {M.~E.}\ \bibnamefont
  {Cates}}\ and\ \bibinfo {author} {\bibfnamefont {J.}~\bibnamefont
  {Tailleur}},\ }\href
  {https://doi.org/10.1146/annurev-conmatphys-031214-014710} {\bibfield
  {journal} {\bibinfo  {journal} {Annual Review of Condensed Matter Physics}\
  }\textbf {\bibinfo {volume} {6}},\ \bibinfo {pages} {219} (\bibinfo {year}
  {2015})}\BibitemShut {NoStop}%
\bibitem [{\citenamefont {Solon}\ \emph {et~al.}(2018)\citenamefont {Solon},
  \citenamefont {Stenhammar}, \citenamefont {Cates}, \citenamefont {Kafri},\
  and\ \citenamefont {Tailleur}}]{solon2018binodal}%
  \BibitemOpen
  \bibfield  {author} {\bibinfo {author} {\bibfnamefont {A.~P.}\ \bibnamefont
  {Solon}}, \bibinfo {author} {\bibfnamefont {J.}~\bibnamefont {Stenhammar}},
  \bibinfo {author} {\bibfnamefont {M.~E.}\ \bibnamefont {Cates}}, \bibinfo
  {author} {\bibfnamefont {Y.}~\bibnamefont {Kafri}},\ and\ \bibinfo {author}
  {\bibfnamefont {J.}~\bibnamefont {Tailleur}},\ }\href
  {https://doi.org/10.1088/1367-2630/aaccdd} {\bibfield  {journal} {\bibinfo
  {journal} {New Journal of Physics}\ }\textbf {\bibinfo {volume} {20}},\
  \bibinfo {pages} {075001} (\bibinfo {year} {2018})}\BibitemShut {NoStop}%
\bibitem [{\citenamefont {Speck}(2022)}]{speck_critical_2022}%
  \BibitemOpen
  \bibfield  {author} {\bibinfo {author} {\bibfnamefont {T.}~\bibnamefont
  {Speck}},\ }\href {https://doi.org/10.1103/PhysRevE.105.064601} {\bibfield
  {journal} {\bibinfo  {journal} {Physical Review E}\ }\textbf {\bibinfo
  {volume} {105}},\ \bibinfo {pages} {064601} (\bibinfo {year}
  {2022})}\BibitemShut {NoStop}%
\bibitem [{\citenamefont {E}\ \emph {et~al.}(2004)\citenamefont {E},
  \citenamefont {Ren},\ and\ \citenamefont {Vanden-Eijnden}}]{weinan2004}%
  \BibitemOpen
  \bibfield  {author} {\bibinfo {author} {\bibfnamefont {W.}~\bibnamefont {E}},
  \bibinfo {author} {\bibfnamefont {W.}~\bibnamefont {Ren}},\ and\ \bibinfo
  {author} {\bibfnamefont {E.}~\bibnamefont {Vanden-Eijnden}},\ }\href@noop {}
  {\bibfield  {journal} {\bibinfo  {journal} {Communications on pure and
  applied mathematics}\ }\textbf {\bibinfo {volume} {57}},\ \bibinfo {pages}
  {637} (\bibinfo {year} {2004})}\BibitemShut {NoStop}%
\bibitem [{\citenamefont {Vanden-Eijnden}\ and\ \citenamefont
  {Heymann}(2008)}]{VE2008gMAM}%
  \BibitemOpen
  \bibfield  {author} {\bibinfo {author} {\bibfnamefont {E.}~\bibnamefont
  {Vanden-Eijnden}}\ and\ \bibinfo {author} {\bibfnamefont {M.}~\bibnamefont
  {Heymann}},\ }\href {https://doi.org/10.1063/1.2833040} {\bibfield  {journal}
  {\bibinfo  {journal} {The Journal of Chemical Physics}\ }\textbf {\bibinfo
  {volume} {128}},\ \bibinfo {pages} {061103} (\bibinfo {year}
  {2008})}\BibitemShut {NoStop}%
\bibitem [{\citenamefont {Heymann}\ and\ \citenamefont
  {Vanden-Eijnden}(2008)}]{heymann2008prl}%
  \BibitemOpen
  \bibfield  {author} {\bibinfo {author} {\bibfnamefont {M.}~\bibnamefont
  {Heymann}}\ and\ \bibinfo {author} {\bibfnamefont {E.}~\bibnamefont
  {Vanden-Eijnden}},\ }\href {https://doi.org/10.1103/PhysRevLett.100.140601}
  {\bibfield  {journal} {\bibinfo  {journal} {Phys. Rev. Lett.}\ }\textbf
  {\bibinfo {volume} {100}},\ \bibinfo {pages} {140601} (\bibinfo {year}
  {2008})}\BibitemShut {NoStop}%
\bibitem [{\citenamefont {Raissi}\ \emph {et~al.}(2019)\citenamefont {Raissi},
  \citenamefont {Perdikaris},\ and\ \citenamefont {Karniadakis}}]{raissi2019}%
  \BibitemOpen
  \bibfield  {author} {\bibinfo {author} {\bibfnamefont {M.}~\bibnamefont
  {Raissi}}, \bibinfo {author} {\bibfnamefont {P.}~\bibnamefont {Perdikaris}},\
  and\ \bibinfo {author} {\bibfnamefont {G.~E.}\ \bibnamefont {Karniadakis}},\
  }\href {https://doi.org/10.1016/j.jcp.2018.10.045} {\bibfield  {journal}
  {\bibinfo  {journal} {Journal of Computational Physics}\ }\textbf {\bibinfo
  {volume} {378}},\ \bibinfo {pages} {686} (\bibinfo {year}
  {2019})}\BibitemShut {NoStop}%
\bibitem [{\citenamefont {Karniadakis}\ \emph {et~al.}(2021)\citenamefont
  {Karniadakis}, \citenamefont {Kevrekidis}, \citenamefont {Lu}, \citenamefont
  {Perdikaris}, \citenamefont {Wang},\ and\ \citenamefont
  {Yang}}]{karniadakis2021}%
  \BibitemOpen
  \bibfield  {author} {\bibinfo {author} {\bibfnamefont {G.~E.}\ \bibnamefont
  {Karniadakis}}, \bibinfo {author} {\bibfnamefont {I.~G.}\ \bibnamefont
  {Kevrekidis}}, \bibinfo {author} {\bibfnamefont {L.}~\bibnamefont {Lu}},
  \bibinfo {author} {\bibfnamefont {P.}~\bibnamefont {Perdikaris}}, \bibinfo
  {author} {\bibfnamefont {S.}~\bibnamefont {Wang}},\ and\ \bibinfo {author}
  {\bibfnamefont {L.}~\bibnamefont {Yang}},\ }\href
  {https://doi.org/10.1038/s42254-021-00314-5} {\bibfield  {journal} {\bibinfo
  {journal} {Nature Reviews Physics}\ }\textbf {\bibinfo {volume} {3}},\
  \bibinfo {pages} {422} (\bibinfo {year} {2021})}\BibitemShut {NoStop}%
\bibitem [{\citenamefont {Simonnet}(2023)}]{simonnet_computing_2023}%
  \BibitemOpen
  \bibfield  {author} {\bibinfo {author} {\bibfnamefont {E.}~\bibnamefont
  {Simonnet}},\ }\href {https://doi.org/10.1016/j.jcp.2023.112349} {\bibfield
  {journal} {\bibinfo  {journal} {Journal of Computational Physics}\ }\textbf
  {\bibinfo {volume} {491}},\ \bibinfo {pages} {112349} (\bibinfo {year}
  {2023})}\BibitemShut {NoStop}%
\bibitem [{\citenamefont {Tjhung}\ \emph {et~al.}(2018)\citenamefont {Tjhung},
  \citenamefont {Nardini},\ and\ \citenamefont {Cates}}]{tjhung_cluster_2018}%
  \BibitemOpen
  \bibfield  {author} {\bibinfo {author} {\bibfnamefont {E.}~\bibnamefont
  {Tjhung}}, \bibinfo {author} {\bibfnamefont {C.}~\bibnamefont {Nardini}},\
  and\ \bibinfo {author} {\bibfnamefont {M.~E.}\ \bibnamefont {Cates}},\ }\href
  {https://doi.org/10.1103/PhysRevX.8.031080} {\bibfield  {journal} {\bibinfo
  {journal} {Physical Review X}\ }\textbf {\bibinfo {volume} {8}},\ \bibinfo
  {pages} {031080} (\bibinfo {year} {2018})}\BibitemShut {NoStop}%
\bibitem [{\citenamefont {Grafke}\ \emph
  {et~al.}(2017{\natexlab{b}})\citenamefont {Grafke}, \citenamefont {Cates},\
  and\ \citenamefont {Vanden-Eijnden}}]{grafke_cates2017}%
  \BibitemOpen
  \bibfield  {author} {\bibinfo {author} {\bibfnamefont {T.}~\bibnamefont
  {Grafke}}, \bibinfo {author} {\bibfnamefont {M.~E.}\ \bibnamefont {Cates}},\
  and\ \bibinfo {author} {\bibfnamefont {E.}~\bibnamefont {Vanden-Eijnden}},\
  }\href {https://doi.org/10.1103/PhysRevLett.119.188003} {\bibfield  {journal}
  {\bibinfo  {journal} {Physical Review Letters}\ }\textbf {\bibinfo {volume}
  {119}},\ \bibinfo {pages} {188003} (\bibinfo {year}
  {2017}{\natexlab{b}})}\BibitemShut {NoStop}%
\bibitem [{\citenamefont {O’Byrne}\ and\ \citenamefont
  {Tailleur}(2020)}]{obyrne2020}%
  \BibitemOpen
  \bibfield  {author} {\bibinfo {author} {\bibfnamefont {J.}~\bibnamefont
  {O’Byrne}}\ and\ \bibinfo {author} {\bibfnamefont {J.}~\bibnamefont
  {Tailleur}},\ }\href@noop {} {\bibfield  {journal} {\bibinfo  {journal}
  {Physical Review Letters}\ }\textbf {\bibinfo {volume} {125}},\ \bibinfo
  {pages} {208003} (\bibinfo {year} {2020})}\BibitemShut {NoStop}%
\bibitem [{\citenamefont {Zakine}\ and\ \citenamefont
  {Vanden-Eijnden}(2023)}]{zakine2022}%
  \BibitemOpen
  \bibfield  {author} {\bibinfo {author} {\bibfnamefont {R.}~\bibnamefont
  {Zakine}}\ and\ \bibinfo {author} {\bibfnamefont {E.}~\bibnamefont
  {Vanden-Eijnden}},\ }\href {https://doi.org/10.1103/PhysRevX.13.041044}
  {\bibfield  {journal} {\bibinfo  {journal} {Phys. Rev. X}\ }\textbf {\bibinfo
  {volume} {13}},\ \bibinfo {pages} {041044} (\bibinfo {year}
  {2023})}\BibitemShut {NoStop}%
\bibitem [{Note1()}]{Note1}%
  \BibitemOpen
  \bibinfo {note} {The Supplemental Material can be found at [INSERT LINK] and
  includes additional references~\cite
  {donnelly_symplectic_2005,forest_fourth_order_1990,ERenVE2002, Cai2, adam}.
  The codes can be found at \protect \url
  {https://github.com/rzakine/gMAM_active-Model-B}.}\BibitemShut {Stop}%
\bibitem [{\citenamefont {Bates}\ and\ \citenamefont
  {Fusco}(2000)}]{bates2000}%
  \BibitemOpen
  \bibfield  {author} {\bibinfo {author} {\bibfnamefont {P.~W.}\ \bibnamefont
  {Bates}}\ and\ \bibinfo {author} {\bibfnamefont {G.}~\bibnamefont {Fusco}},\
  }\href {https://doi.org/10.1006/jdeq.1999.3660} {\bibfield  {journal}
  {\bibinfo  {journal} {Journal of Differential Equations}\ }\textbf {\bibinfo
  {volume} {160}},\ \bibinfo {pages} {283} (\bibinfo {year}
  {2000})}\BibitemShut {NoStop}%
\bibitem [{\citenamefont {Bates}\ and\ \citenamefont {Fife}(1993)}]{bates1993}%
  \BibitemOpen
  \bibfield  {author} {\bibinfo {author} {\bibfnamefont {P.~W.}\ \bibnamefont
  {Bates}}\ and\ \bibinfo {author} {\bibfnamefont {P.~C.}\ \bibnamefont
  {Fife}},\ }\href {https://doi.org/10.1137/0153049} {\bibfield  {journal}
  {\bibinfo  {journal} {SIAM Journal on Applied Mathematics}\ }\textbf
  {\bibinfo {volume} {53}},\ \bibinfo {pages} {990} (\bibinfo {year}
  {1993})}\BibitemShut {NoStop}%
\bibitem [{\citenamefont {Donnelly}\ and\ \citenamefont
  {Rogers}(2005)}]{donnelly_symplectic_2005}%
  \BibitemOpen
  \bibfield  {author} {\bibinfo {author} {\bibfnamefont {D.}~\bibnamefont
  {Donnelly}}\ and\ \bibinfo {author} {\bibfnamefont {E.}~\bibnamefont
  {Rogers}},\ }\href {https://doi.org/10.1119/1.2034523} {\bibfield  {journal}
  {\bibinfo  {journal} {American Journal of Physics}\ }\textbf {\bibinfo
  {volume} {73}},\ \bibinfo {pages} {938} (\bibinfo {year} {2005})}\BibitemShut
  {NoStop}%
\bibitem [{\citenamefont {Forest}\ and\ \citenamefont
  {Ruth}(1990)}]{forest_fourth_order_1990}%
  \BibitemOpen
  \bibfield  {author} {\bibinfo {author} {\bibfnamefont {E.}~\bibnamefont
  {Forest}}\ and\ \bibinfo {author} {\bibfnamefont {R.~D.}\ \bibnamefont
  {Ruth}},\ }\href {https://doi.org/10.1016/0167-2789(90)90019-L} {\bibfield
  {journal} {\bibinfo  {journal} {Physica D: Nonlinear Phenomena}\ }\textbf
  {\bibinfo {volume} {43}},\ \bibinfo {pages} {105} (\bibinfo {year}
  {1990})}\BibitemShut {NoStop}%
\bibitem [{\citenamefont {E}\ \emph {et~al.}(2002)\citenamefont {E},
  \citenamefont {Ren},\ and\ \citenamefont {Vanden-Eijnden}}]{ERenVE2002}%
  \BibitemOpen
  \bibfield  {author} {\bibinfo {author} {\bibfnamefont {W.}~\bibnamefont {E}},
  \bibinfo {author} {\bibfnamefont {W.}~\bibnamefont {Ren}},\ and\ \bibinfo
  {author} {\bibfnamefont {E.}~\bibnamefont {Vanden-Eijnden}},\ }\href
  {https://doi.org/10.1103/PhysRevB.66.052301} {\bibfield  {journal} {\bibinfo
  {journal} {Phys. Rev. B}\ }\textbf {\bibinfo {volume} {66}},\ \bibinfo
  {pages} {052301} (\bibinfo {year} {2002})}\BibitemShut {NoStop}%
\bibitem [{\citenamefont {Liu}\ \emph {et~al.}(2020)\citenamefont {Liu},
  \citenamefont {Cai},\ and\ \citenamefont {Xu}}]{Cai2}%
  \BibitemOpen
  \bibfield  {author} {\bibinfo {author} {\bibfnamefont {Z.}~\bibnamefont
  {Liu}}, \bibinfo {author} {\bibfnamefont {W.}~\bibnamefont {Cai}},\ and\
  \bibinfo {author} {\bibfnamefont {J.~Z.-Q.}\ \bibnamefont {Xu}},\ }\href
  {https://doi.org/https://doi.org/10.4208/cicp.OA-2020-0179} {\bibfield
  {journal} {\bibinfo  {journal} {Communications in Computational Physics}\
  }\textbf {\bibinfo {volume} {28}},\ \bibinfo {pages} {1970} (\bibinfo {year}
  {2020})}\BibitemShut {NoStop}%
\bibitem [{\citenamefont {Kingma}\ and\ \citenamefont {Ba}(2014)}]{adam}%
  \BibitemOpen
  \bibfield  {author} {\bibinfo {author} {\bibfnamefont {D.}~\bibnamefont
  {Kingma}}\ and\ \bibinfo {author} {\bibfnamefont {J.}~\bibnamefont {Ba}},\
  }\href@noop {} {\bibfield  {journal} {\bibinfo  {journal} {arXiv:1412.6980}\
  } (\bibinfo {year} {2014})}\BibitemShut {NoStop}%
\end{thebibliography}%

\clearpage


\section*{Supplementary material}

# Supplemental Material: Unveiling the Phase Diagram and Reaction Paths of the Active Model B with the Deep Minimum Action Method

Ruben Zakine
*Courant Institute, New York University, 251 Mercer Street, New York, New York 10012, USA*
*Chair of Econophysics and Complex Systems, École Polytechnique, 91128 Palaiseau Cedex, France and*
*LadHyX UMR CNRS 7646, École Polytechnique, 91128 Palaiseau Cedex, France*

Éric Simonnet
*INPHYNI, Universitée Côte d'Azur et CNRS, UMR 7010, 1361, route des Lucioles, 06560, Valbonne, France*

Eric Vanden-Eijnden
*Courant Institute, New York University, 251 Mercer Street, New York, New York 10012, USA*
(Dated: June 28, 2024)

## I. COMPUTING THE CRITICAL STATES

To compute the critical states we must calculate the fixed points of the noiseless equation:

$$\partial_t \phi = -\partial_x^2[\nu \partial_x^2 \phi + \phi - \phi^3 - \lambda(\partial_x \phi)^2]. \tag{1}$$

Since space can be rescaled, here we work on a domain of size $L = 1$ where $\nu$ can vary (in the main text, we rather chose to work with $\nu = 1$ and varying $L$). Using stationarity and the periodic boundary conditions, a critical points $\phi_c$ must satisfy

$$\nu \partial_x^2 \phi_c + \phi_c - \phi_c^3 - \lambda(\partial_x \phi_c)^2 - \mu_0 = 0, \tag{2}$$

with $\mu_0$ a constant, and $\phi_c$ subject to constraints of periodicity and $\int_0^1 \phi_c(x)dx = \phi_0$. The Newton mapping rephrases the stationary equation into an equation of motion, see e.g. [1]. With the correspondence $x \leftrightarrow t$ and $\phi_c \leftrightarrow z$, one gets

$$\nu \ddot{z}(t) = -z(t) + z^3(t) + \lambda(\dot{z}(t))^2 + \mu_0, \tag{3}$$

where the $\dot{z} \equiv dz/dt$.

It turns out that this system is integrable, whether $\lambda = 0$ or not. Indeed, one can set $v(z(t)) \equiv \dot{z}$, such that $\ddot{z} = \frac{dv}{dz}\frac{dz}{dt} = \frac{dv}{dz}v$, which enables us to write

$$\begin{cases} \dot{z} = v \\ \dfrac{d}{dz}(v^2) - \dfrac{2\lambda}{\nu}v^2 = \dfrac{2(\mu_0 - z + z^3)}{\nu}, \end{cases} \tag{4}$$

whose solution takes the form

$$v^2(z) = e^{2\alpha z}\left(K + \int^z h(z')e^{-2\alpha z'}dz'\right), \tag{5}$$

with $\alpha = \lambda/\nu$, $h(z) = 2(\mu_0 - z + z^3)/\nu$, and $K$ some constant. For $\alpha = 0$, eq. (5) simplifies into

$$v^2 = K - \frac{2}{\nu}V(z), \tag{6}$$

with $V(z) = z^2/2 - z^4/4 - \mu_0 z$. For $\alpha \neq 0$, one computes explicitly

$$e^{2\alpha z}\int^z h(z')e^{-2\alpha z'}dz' \tag{7}$$

$$= -\frac{3 + 6\alpha z + 2(3z^2 - 1)\alpha^2 + 4(\mu_0 - z + z^3)\alpha^3}{4\alpha^4 \nu} \tag{8}$$

$$\equiv g(z), \tag{9}$$

such that we have

$$v^2 = e^{2\alpha z}K + g(z). \tag{10}$$

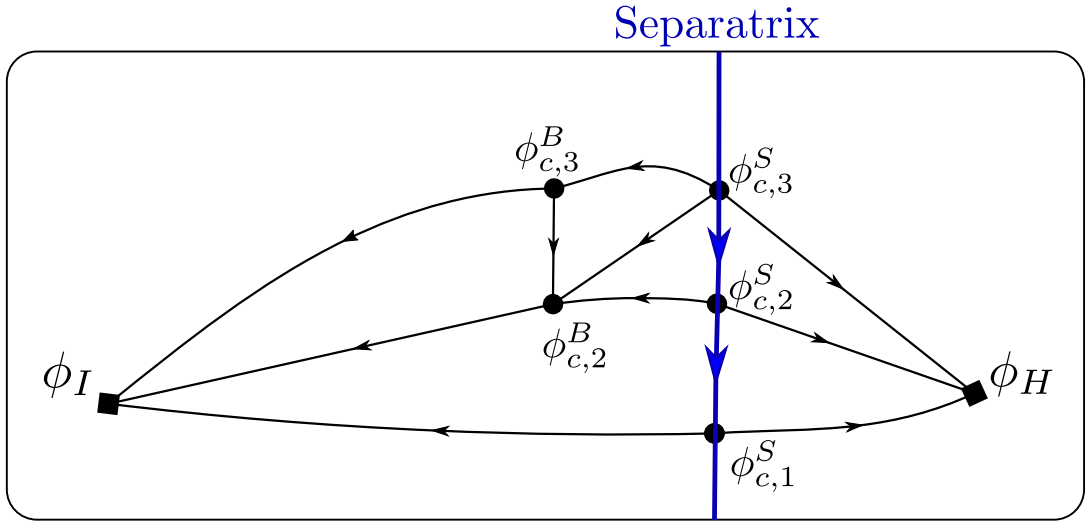

FIG. 1. Sketch of the relaxation flow lines joining different critical points with number of bumps $q < 4$. A directed link corresponds to an existing relaxation path between two critical points (the arrow indicates the noiseless flow). Critical states $\phi_{c,q}^{S,B}$ appear as pairs in a saddle-node bifurcation. States $\phi_{c,q}^S$ lie on the separatrix, states $\phi_{c,q}^B$ lie in the basin of attraction of $\phi_I$. The separatrix is a high-dimensional manifold that cannot be easily represented. The path are obtained with the string method, and the action is computed along each path to check that it is indeed 0 along a relaxation path.

Since a conserved quantity exists whether $\lambda = 0$ or not, one should resort to a symplectic scheme [2] to find the periodic orbits that need to be computed with high accuracy. We use the high-order symplectic scheme proposed by Forest and Ruth [3]. For $\lambda = 0$, the equation of motion is simply

$$\nu \ddot{z} = -z + z^3 + \mu_0, \tag{11}$$

and for $\lambda \neq 0$ we instead solve:

$$\ddot{z} = \alpha e^{2\alpha z} K + \frac{g'(z)}{2} \tag{12}$$

$$= \alpha e^{2\alpha z} K - \frac{3 + 6\alpha z + \alpha^2(6z^2 - 2)}{4\alpha^3 \nu}, \tag{13}$$

where $K = -g(z_0)e^{-2\alpha z_0}$ is obtained from (10) considering the initial conditions $(z(0), \dot{z}(0)) = (z_0, 0)$.

The orbits are closed and $z(t)$ evolves between the maximum and the minimum root of $-V'(X) = X^3 - X + \mu_0$. This notably constrains $|\mu_0| < 2\sqrt{3}/9$ for $V'(X)$ to display 3 real roots. Denoting $r_1, r_2, r_3$ (with $r_1 < r_2 < r_3$) the roots of $V'(X)$, we initialize the dynamics close to $r_3$, for instance $(z(0), \dot{z}(0)) = (r_3 - \delta, 0)$, with $\delta \ll 1$. We run the symplectic scheme and compute the distance between the position at time 1 and the position at time 0, $|z(1) - z(0)| \equiv \xi$. The parameter $\delta$ is varied until we obtain $\xi < 10^{-8}$, which is our criterion to obtain periodic orbits. Note that for given $(\nu, \alpha, \mu_0)$, there may be several values $\delta_1 < \delta_2 < \cdots < \delta_q$ each leading to different periodic orbits $\phi_{c,1}, \phi_{c,2}, \ldots, \phi_{c,q}$, that each display a different number $q$ of bumps. The symplectic scheme is implemented with $\Delta t = 10^{-4}$, such that the number of points on the orbit is $10^4$. For each acceptable periodic orbit, we compute $\int_0^1 z(t)dt \equiv \bar{z}_q(\nu, \alpha, \mu_0)$ where the index $q$ indicates the number of periods on the domain $[0, 1]$. The equation

$$\phi_0 - \bar{z}_q(\nu, \alpha, \mu_0) = 0, \tag{14}$$

where $\mu_0$ is the unknown, has always 0, 1 or 2 solutions. When no solution exists, it means that no critical states with a given number $q$ of bumps are accessible. When two solutions exist for $\mu_0$, it means that two periodic orbits of mass $\phi_0$ can be found for given $(\nu, \alpha)$. These two orbits correspond to two critical states (see e.g. Fig. 2 in main text), one state sitting on the separatrix between $\phi_I$ and $\phi_H$, and the other lying strictly in the basin of attraction of $\phi_I$ (see Fig. 1 in Appendix, and see main text). When (14) has one solution only, then there is a unique critical orbit of $q$ bumps and mass $\phi_0$ accessible. This critical state is denoted $\phi_{c,q}^\star$ in the main text. Going from zero to two solutions in (14) corresponds to a saddle-node bifurcation. For fixed density $\phi_0$, the manifolds on which the saddle-node bifurcations occur for a given $q$ are displayed in Fig. 2c in the main text.

## II. DETAILS ON THE GMAM IMPLEMENTATION

In this section we detail the numerical algorithm based on the classic gMAM [4] that we run to obtain the minimum action path. Before running the gMAM, we first run the string method detailed in [5]. The starting point of the string method is to consider a path $\{\varphi_i\}$ made of $N$ successive states interpolating between $\phi_H$ and $\phi_I$. We set $\varphi_1 = \phi_H(x) \equiv \phi_0$ and $\varphi_N = \phi_I(x)$, but since $\phi_I(x)$ is not known analytically, it is obtained from the noiseless evolution of $\phi$, starting from any state in the basin of attraction of $\phi_I$. To solve the evolution equation of $\phi(x,t)$, space is discretized on a 1-d domain with periodic boundary conditions. We use a semi-implicit and pseudo-spectral scheme: the $\partial_x^4 \phi$ term is treated implicitly in time in Fourier space and all the other terms are treated explicitly in time in real space.

Now, at each iteration of the string method, we perform two steps: (i) for all $i \in \{1, \ldots, N\}$, the states $\varphi_i$ along the path perform a relaxation step using our semi-implicit scheme; (ii) after the relaxation step, a new path with regularly spaced states replaces the former path. The states along the new path are obtained by interpolation and arclength renormalization. This step prevents the states to collapse onto $\phi_I$ or $\phi_H$, as they would do if only the relaxation was implemented. The endpoints of the string, $\varphi_1$ and $\varphi_N$, are left free and keep undergoing pure relaxation evolution to ensure that the two fixed points $\phi_I$ and $\phi_H$ are indeed dynamically stable. At the end of the string algorithm, one obtains the heteroclinic orbit joining $\phi_I$ to $\phi_H$ and going through an unstable fixed point of Morse index 1. The heteroclinic orbit (or relaxation path) is then a good candidate for the instanton (since it is indeed the instanton in equilibrium).

We then run the gMAM algorithm with the string as a starting point. The geometric implementation allows for reaching reaction paths of infinite duration. We follow a hybrid implementation of the gMAM, using both the Lagrangian formulation [4, 6] and the Hamiltonian formulation to update the Lagrange multiplier [7]. We recall that in the presence of a conserved Gaussian white noise, the action we consider reads

$$S_T = \int_0^T \|\partial_t \phi - F_{\text{AMB}}(\phi)\|_{-1}^2 dt, \tag{15}$$

with the flow of the active Model B given by $F_{\text{AMB}}(\phi) = \partial_x^2(-\nu \partial_x^2 \phi + \phi^3 - \phi + \lambda(\partial_x \phi)^2)$, and where the norm $\|\cdot\|_{-1}$ is defined from a scalar product $\langle \cdot, \cdot \rangle_{-1}$ on any spatial domain $\Omega$:

$$\langle u, v \rangle_{-1} = \int_\Omega u(-\Delta^{-1} v), \tag{16}$$

with $\Delta^{-1} \equiv \partial_x^{-2}$ the inverse Laplacian. The minimum action method performs a gradient descent on the action $S_T$ to obtain the instantons. Introducing an artificial time $\tau$ to evolve the path $\{\phi(t,x)\}_{t \in [0,T]}$, one solves numerically (see Ref. [6])

$$\partial_\tau \phi(t, x; \tau) = -\frac{\delta S_T}{\delta \phi(t, x)}. \tag{17}$$

When $\phi$ eventually solves $\frac{\delta S_T}{\delta \phi} = 0$, the Euler-Lagrange equation is satisfied and $\{\phi(t, x)\}_{t \in [0,T]}$ is the instanton. For convenience, we work with $\theta$, the conjugate field of $\phi$ in the Hamiltonian formulation, which solves the Hamilton equations of motion for the instanton. Here the Hamiltonian explicitly reads

$$H(\phi, \theta) = \int_\Omega \left[ F_{\text{AMB}}(\phi) \theta + \frac{1}{2}(\partial_x \theta)^2 \right] dx, \tag{18}$$

and the instanton solves (in addition to the boundary conditions at $t = 0, T$)

$$\begin{cases} \partial_t \theta = -\partial_\phi H(\phi, \theta) \\ \partial_t \phi = \partial_\theta H(\phi, \theta). \end{cases} \tag{19}$$

or, explicitly for the second equation,

$$\partial_t \phi = F_{\text{AMB}}(\phi) - \partial_x^2 \theta. \tag{20}$$

The gradient descent (17) amounts to solving

$$\partial_\tau \phi = \partial_t \theta + \frac{\delta H}{\delta \phi}. \tag{21}$$

Using the explicit formula for $\theta$ from (20), namely

$$\theta = -\Delta^{-1}(\partial_t \phi - F_{\text{AMB}}(\phi)), \tag{22}$$

---

**Algorithm 1** : Geometric Minimum Action Method, pseudo-spectral and semi-implicit scheme

1: **Inputs:** $N \in \mathbb{N}$; two stable fixed points $\phi_a$ and $\phi_b$ of the noiseless dynamics; a path $\{\hat{\phi}_i^0\}_{i \in I}$ and $I = \{1, \cdots, N\}$, with $\hat{\phi}_0^0 = \phi_a$ and $\hat{\phi}_M^0 = \phi_b$, such that $|\hat{\phi}_{i+1}^0 - \hat{\phi}_i^0|$ is constant in $i$; the functions $F_{\text{AMB}}(\phi), \gamma(\phi, \theta); \Delta\tau > 0$,
2: **Initialization:** For every $i \in I$, take $\hat{\theta}_i^0 = 0$; set $\Delta s = 1/N$.
3: **for** $n \geq 0$ **do**
4:     Compute $F_{\text{AMB}}(\phi^n)$ in real space.
5:     Compute $\gamma^n = \gamma(\hat{\phi}^n, \hat{\theta}^n)$.
6:     Compute the Fourier transform $\mathcal{F}_k(F_{\text{AMB}})$.
7:     Compute in Fourier space $\bar{\theta}_{i,k}^n$:

$$\bar{\theta}_{i,k}^n = k^{-2} \mathcal{F}_k(\gamma^n \frac{\hat{\phi}_{i+1}^n - \hat{\phi}_i^n}{\Delta\tau} - F_{\text{AMB}}(\phi^n))$$

8:     Update $\phi$ with the semi-implicit Thomas algorithm, namely, solve $\bar{\phi}^{n+1}$

$$\frac{\bar{\phi}_{i,k}^{n+1} - \bar{\phi}_{i,k}^n}{\Delta\tau} = k^{-2}(\gamma_i^n)^2 \frac{\bar{\phi}_{i+1,k}^{n+1} + \bar{\phi}_{i-1,k}^{n+1} - 2\bar{\phi}_{i,k}^{n+1}}{(\Delta\tau)^2} + k^{-2}\gamma_i^n \gamma_i^{n\prime} \bar{\phi}_i^{n\prime} + k^{-2}\gamma_i^n \mathcal{F}_k[F_{\text{AMB}}(\phi^n)]_i'$$
$$+ \mathcal{F}_k \left( \left\langle \frac{\delta F_{\text{AMB}}(\hat{\phi}^n)}{\delta \hat{\phi}^n}, \gamma_i^n \hat{\phi}_i^{n\prime} \right\rangle_{-1} \right) - \mathcal{F}_k \left( \left[ \left\langle \frac{\delta F_{\text{AMB}}(\hat{\phi}^n)}{\delta \hat{\phi}^n}, F_{\text{AMB}}(\hat{\phi}^n) \right\rangle_{-1} \right]_i' \right).$$

    For the arclength derivative of a function $g$, we take $g_i' = (g_{i+1} - g_i)/(\Delta s)$ for $i \in [1, N-1]$, and $g_M' = (g_N - g_{N-1})/(\Delta s)$ for the endpoint.
9:     Compute the inverse Fourier transform of $(\bar{\phi}^{n+1}, \bar{\theta}^n) = (\mathcal{F}_k^{-1}\bar{\phi}_k^n, \mathcal{F}_k^{-1}\bar{\theta}_k^n)$.
10:     Interpolate $\{(\bar{\phi}_i^{n+1}, \bar{\theta}_i^{n+1})\}_{i \in I}$ onto a path $\{(\hat{\phi}_i^{n+1}, \hat{\theta}_i^{n+1})\}_{i \in I}$ such that $|\hat{\phi}_{i+1}^{n+1} - \hat{\phi}_i^{n+1}|$ is constant in $i$, as in the string method.

---

and using

$$\frac{\delta H}{\delta \phi} = \left\langle \frac{\delta F_{\text{AMB}}(\phi)}{\delta \phi}, \partial_t \phi - F_{\text{AMB}}(\phi) \right\rangle_{-1}, \tag{23}$$

one obtains

$$\partial_\tau \phi = -\Delta^{-1} \partial_t^2 \phi + \partial_t \Delta^{-1} F_{\text{AMB}}(\phi) + \left\langle \frac{\delta F_{\text{AMB}}(\phi)}{\delta \phi}, \partial_t \phi \right\rangle_{-1}$$
$$- \left\langle \frac{\delta F_{\text{AMB}}(\phi)}{\delta \phi}, F_{\text{AMB}}(\phi) \right\rangle_{-1}. \tag{24}$$

This equation can be solved with a pseudo-spectral method since the operator $-\Delta^{-1}$ is diagonal in Fourier space. The relaxation along $t$ contains a diffusion term that can be treated implicitly by means of a Thomas algorithm [4, 6]. In addition, the flow $F_{\text{AMB}}(\phi)$ can further decomposed into the

$$F_{\text{AMB}}(\phi) = -\nu \Delta^2 \phi + \mathcal{K}\phi, \tag{25}$$

with $\mathcal{K}\phi = \partial_x^2(\phi^3 - \phi + \lambda(\partial_x \phi)^2)$, such that the relaxation of the field in $x$ can also be treated implicitly in $\tau$ when diagonalizing the operator $\Delta^2$. Without this decomposition, the Courant–Friedrichs–Lewy condition for the scheme to be stable is extremely restrictive.

Finally, we use arclength parametrization of the path such that time $t$ is a function of arclength $s \in [0, 1]$, and instantons between fixed points that require $t \to \infty$ can then be computed. In particular, we now have $\partial_t \phi(t) = (dt/ds)^{-1} \partial_s \hat{\phi}(s) \equiv \gamma(s)\hat{\phi}'$, and $\partial_t \theta(t) = (dt/ds)^{-1} \partial_s \hat{\theta}(s) \equiv \gamma(s)\hat{\theta}'$, where the $'$ denotes the derivative w.r.t arclength $s$. The function $\gamma(s)$ is also a Lagrange multiplier enforcing $H = 0$ along the instanton [4]. The geometric minimization now reads

$$\partial_\tau \hat{\phi} = -\Delta^{-1} \gamma^2 \partial_s^2 \hat{\phi} - \Delta^{-1} \gamma \gamma' \partial_s \hat{\phi} + \gamma \partial_s \Delta^{-1} F_{\text{AMB}}(\hat{\phi})$$
$$+ \left\langle \frac{\delta F_{\text{AMB}}(\hat{\phi})}{\delta \hat{\phi}}, \gamma \partial_s \phi \right\rangle_{-1} - \left\langle \frac{\delta F_{\text{AMB}}(\hat{\phi})}{\delta \hat{\phi}}, F_{\text{AMB}}(\hat{\phi}) \right\rangle_{-1}. \tag{26}$$

Note that $\gamma(s)$ is numerically computed using [7]:

$$\gamma = \frac{|\langle \partial_\theta H, \partial_s \hat{\phi} \rangle| + \sqrt{\max(0, \langle \partial_\theta H, \partial_s \hat{\phi} \rangle^2 - 4H|\partial_s \hat{\phi}|^2)}}{2|\partial_s \hat{\phi}|^2}. \tag{27}$$

The complete algorithm is summarized in Algorithm 1. It is worth mentioning that the equation, already in 1d, is very ill-conditioned, and forces us to take a fine grid, which in turn leads to very long computation times. Typically, for physical parameters $L = 1$, $\nu = 5 \times 10^{-4}$, $\alpha = 2$, $\phi_0 = 0.65$, one must take $N_x = 256$ space grid points, $N_s = 400$ arclength grid, and $\Delta \tau = 10^{-6}$ to reach the minimum action path in a typical computation time of $\simeq 3$ weeks on a 2.8GHz machine.

### III. DEEP GMAM METHOD: COMPUTING REACTION PATHS USING NEURAL NETWORKS

Here we detail the deep gMAM algorithm [8]. This method differs from the classical gMAM algorithm in two significant ways. Firstly, instead of decomposing the path into ad-hoc space-time bases (e.g., pseudo-spectral decomposition, finite differences method, etc...), the reaction path is parametrized in both space and time by a neural network. Secondly, it directly tackles the geometric action as the cost functional to minimize, as opposed to finding the zeros of the Euler-Lagrange equation. There are several advantages to this approach. The remarkable expressivity of neural networks and their ability to overcome the curse of dimensionality enable the consideration of higher dimensional problems. Furthermore, the method determines its own parametrization rather than relying on a rigid constant arclength constraint. This method also offers high flexibility, providing multiple approaches to address the minimization problem and construct ad-hoc surrogate/ansatz for the reaction path. However, there are some drawbacks. Tuning of the hyperparameters is often necessary, and the precision may be degraded compared to classical methods, particularly when dealing with ill-conditioned problems.

Below is a detailed description of the algorithm used to obtain the various reaction paths and action values in Figures 1, 3 and 4 in the main text. Let $d$ be the spatial dimension, where $d = 1$ or $d = 2$ for Figure 4. We define $u : (s, \mathbf{x}) \in (0, 1) \times (0, 1)^d \mapsto u(s, \mathbf{x})$ some reaction path such that $u(0, \mathbf{x}) = a(\mathbf{x})$ and $u(1, \mathbf{x}) = b(\mathbf{x})$, where $a, b$ are two given fields, for instance $a = \phi_I$ (inhomogeneous state) and $b = \phi_H$ (homogeneous state).

To minimize the AMB geometric action, four important constraints must be taken into account:

1. It is a boundary-value problem;

2. The mass of the solution $\phi_0$ must be conserved along the reaction path, namely $\langle u \rangle \equiv \int_\Omega u(s, \mathbf{x}) \, d\mathbf{x} = \phi_0$ for all $s \in (0, 1)$;

3. The noise conservation (see the term $\partial_x \xi$ in Eq.(5)) implies that one must work with respect to the $H^{-1}$ norm;

4. The periodic boundary conditions must also be satisfied.

As shown below, it is possible to formulate the problem in a way that eliminates three of these constraints, leaving only the noise conservation constraint. This remaining constraint must be explicitly enforced through penalization.

We need to find the minimizers of the geometric action:

$$\mathcal{A}_g[u] = \int_0^1 (||\dot{u}||_{-1} ||F(u)||_{-1} - \langle \dot{u}, F(u) \rangle_{-1}) \, ds,$$

where $\langle u, v \rangle_{-1} \equiv \langle u, -\Delta^{-1} v \rangle$, $\dot{u} = du/ds$, and $F(u) = -\Delta F_{\text{aGL}}(u)$, with $F_{\text{aGL}}(u) = \nu \Delta u + u - u^3 - \lambda |\nabla u|^2$, the flow of the active Ginzburg-Landau dynamics. The problem can be reformulated as:

$$\mathcal{A}_g[u] = \int_0^1 (||\nabla v|| \, ||\nabla F_{\text{aGL}}|| - \langle \nabla v, \nabla F_{\text{aGL}} \rangle) \, ds$$

with the constraint

$$-\Delta v = \dot{u}.$$

We now consider a neural network parameterization for $u$. We use the simplest fully connected architecture given by $\mathcal{N}_u(s, \mathbf{x}; \mathcal{T}) = \mathcal{N}_{D+1} \circ \mathcal{N}_D \cdots \mathcal{N}_1 \circ \mathcal{N}_0(s, \mathbf{x})$, where $\mathcal{N}_k(\mathbf{y}) = \sigma_k(W_k \mathbf{y} + \mathbf{b}_k)$ for $k = 0, \cdots, D+1$. Here, $\mathcal{T}$ is the set of the NN parameters weights/bias $\{W_k, \mathbf{b}_k\}$, $\sigma_k$ are the activation functions, and $D$ is often called the depth or number of hidden layers. In our setting, we use swish activation function with $\text{swish}(x) = x/(1 + e^{-x})$ for all $k < D+1$ and a linear output for the last layer $k = D+1$. The matrices $W_k$ have size $c \times c$ for all $k = 1, \cdots, D$, $W_0$ has size $c \times d + 1$, and the

last output layer has size $1 \times c$, where $c$ is the number of neurons per layer, often called capacity. A naive approach would be to directly substitute u with its neural network representation, denoted as $\mathcal{N}_u$. However, this would necessitate penalizing the remaining constraints. Instead, we adopt the following ansatz:

$$\begin{aligned}\mathcal{U}(s,\mathbf{x}) =& (1-s)a(\mathbf{x}) + sb(\mathbf{x}) \\ & + s(1-s)\left(\mathcal{N}_{u,\mathrm{per}}(s,\boldsymbol{\varphi}) - \langle \mathcal{N}_{u,\mathrm{per}}\rangle\right),\end{aligned} \quad (28)$$

with

$$\boldsymbol{\varphi}(\mathbf{x}) \equiv (\cos 2\pi\mathbf{x}, \sin 2\pi\mathbf{x}). \quad (29)$$

It is easy to check that $\mathcal{U}(0,\mathbf{x}) = a(\mathbf{x})$, $\mathcal{U}(1,\mathbf{x}) = b(\mathbf{x})$. Moreover, we have $\langle \mathcal{U}\rangle = (1-s)\langle a\rangle + s\langle b\rangle + 0 = \phi_0$ since $\langle a\rangle = \langle b\rangle = \phi_0$ by hypothesis. Finally, it is periodic on the domain $(0,1)^d$. In fact, $\mathcal{N}_{u,\mathrm{per}}$ depends nonlinearly on $\boldsymbol{\varphi}$, so that adding higher harmonics in the input is unnecessary. Note that the input dimension has increased from $1+d$ to $1+2d$. The term $\langle \mathcal{N}_{u,\mathrm{per}}\rangle$ is numerically approximated on a Gauss-Legendre grid $\{\omega_k, \mathbf{x}_k\}$:

$$\langle \mathcal{N}_{u,\mathrm{per}}\rangle \approx \sum_{k=1}^{N_g} \omega_k \mathcal{N}_{u,\mathrm{per}}(s,\boldsymbol{\varphi}).$$

The noise constraint imposes the need for another periodic NN to represent the function $v$, namely we use the ansatz

$$\mathcal{V}(s,\mathbf{x}) = \mathcal{N}_{v,\mathrm{per}}(s,\boldsymbol{\varphi}),$$

and minimize the residual $||\Delta\mathcal{V} + \dot{\mathcal{U}}||$ for all $s$. Due to the ill-conditioned nature of the problem, it is beneficial to replace both $\mathcal{N}_{u,\mathrm{per}}$ and $\mathcal{N}_{v,\mathrm{per}}$ with more expressive neural network terms. Specifically, each of these periodic NNs is replaced by a sum of $K$ independent NNs with rescaled inputs. For example, the periodic NN for $u$ is given by

$$\mathcal{N}_{u,\mathrm{per}}(s,\boldsymbol{\varphi}) = \sum_{k=0}^{K-1} \mathcal{N}_k(4^k(s,\boldsymbol{\varphi})),$$

where each $\mathcal{N}_k$ represents an independent neural network. This replacement enables the high-frequency terms to be learned more quickly (see e.g. [9]). We set $K=4$ in practice.

The final cost functional that we aim to minimize is given by

$$\mathcal{C}[\mathcal{U},\mathcal{V}] = \mathcal{A}_g[\mathcal{U}] + \gamma_v \int_0^1 ||\Delta\mathcal{V} + \dot{\mathcal{U}}||^2\, ds +$$

$$\gamma_{\mathrm{arc}}\left(\int_0^1 ||\dot{\mathcal{U}}||^2\, ds - \left(\int_0^1 ||\dot{\mathcal{U}}||\, ds\right)^2\right),$$

where $\gamma_v$ and $\gamma_{\mathrm{arc}}$ are penalty coefficients. We use a small penalty coefficient $\gamma_{\mathrm{arc}} \ll 1$ so that constant arclength parametrization is favored but not strictly enforced. On the other hand, $\gamma_v$ must be chosen to be large enough. In practice, we use $\gamma_v = O(10)$ initially and increase it to $O(100)$ towards the end of the simulations.

Once the cost functional is written, the minimization follows classical machine learning techniques. We use here Monte-Carlo batches by sampling uniformly $(s,\mathbf{x}) \sim U([0,1]) \times U([0,1]^d)$. We call $N_s$ the number of points for $s$ and $N_\mathbf{x}$ for the variable $\mathbf{x}$. One must then compute the gradients of the cost functional w.r.t. the NNs parameters $\nabla_{\mathcal{T}_u,\mathcal{T}_v}\mathcal{C}[\mathcal{U},\mathcal{V}](s_i,x_i), i = 1,\cdots N_s \times N_\mathbf{x}$ and use a gradient descent approach (here, ADAM method [10]). The parameters used for Fig. 4 are $c = 15$, $D = 15$, $K = 4$, $N_s = 40$, $N_\mathbf{x} = 120$, $N_g = 30^2$ with swish activation function and learning rate $\eta = 10^{-3}$ which is decreased to $10^{-4}$ at the end of the simulation. The penalization coefficients are $\gamma_v = 10$, $\gamma_{\mathrm{arc}} = 10^{-2}$ and $\gamma_v$ is increased to 100 at the end of the simulation giving a $L^2$ residual for the Poisson constraint $\approx 10^{-3}$. The physical parameters are $\phi_0 = 0.65$, $L = 44.7\nu^{1/2}$, and $\lambda = 2\nu$. We also illustrate the efficiency of the method by displaying an asymmetric path located above the (*) symbol in Figure 3 of the main text, it is shown in Fig. 2. Such asymmetric configurations are always observed when the reaction path exhibits an even number of bumps (here, 2 bumps in Fig. 2). They have lower action values than their symmetric versions suggesting that the breaking of the symmetry $x \to -x$ facilitates transitions.

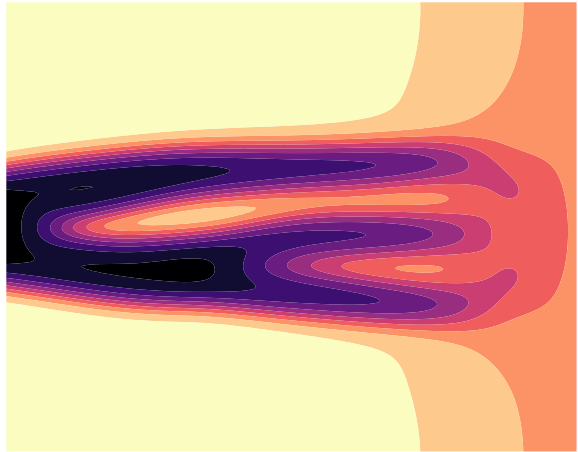

FIG. 2. Asymmetric instanton for $\phi_0 = 0.65$, $L = 50\nu^{1/2}$ and $\alpha = 2$ with lower action value $\approx 0.0215$ compared to its symmetric counterpart with action value $\approx 0.0235$ ($L = 1$).

## IV. COMPUTATION OF AMB CRITICAL POINTS BY NEURAL NETWORKS IN DIMENSION $d \geq 2$

The computation of the inhomogeneous state $\phi_I$ is a prerequisite in the previous section. Additionally, it is often of interest to investigate the reaction paths that connect $\phi_I$ to some critical points, i.e., the ending state $b$ is a chosen critical point that satisfies

$$\nabla F_{\mathrm{aGL}}(u) = 0. \tag{30}$$

In two dimensions, the one-dimensional techniques discussed above do not hold anymore, and one must solve Eq. (30) directly. One expects to have not just one but many solutions as $\nu \to 0$. As in the previous section, two constraints are present: the periodic constraint and the mass constraint $\langle u \rangle = \phi_0$. A convenient NN ansatz that replaces $u$ and satisfies both constraints is given by

$$\mathcal{U}(\mathbf{x}) = f(\mathbf{x}) - \langle f \rangle + \mathcal{N}_u(\boldsymbol{\varphi}) - \langle \mathcal{N}_u \rangle + \phi_0,$$

where $f$ is chosen to initialize the gradient descent properly since at the beginning of the descent, $\mathcal{N}_u$ is typically small, so that $\mathcal{U} \approx f - \langle f \rangle + \phi_0$. One then minimizes the cost functional

$$\mathcal{C}[\mathcal{U}] = ||\nabla F_{\mathrm{aGL}}||^2. \tag{31}$$

We show below that the critical points obtained are spikes [11] constrained by $C_n$ (or product of) discrete rotation symmetries. In order to compute these critical points, we choose $f(x)$ as a sum of exponentials: $f(x) = c_0 - c_1 \sum_{k=1}^{n} e^{-\gamma ||\mathbf{x} - \mathbf{x}_k||^2}$ where $\mathbf{x}_k$ are chosen arbitrarily and $c_0, c_1$ are constants so that $\langle f \rangle = \phi_0$. Due to the torus geometry, purely radial solutions cannot exist. However, since solutions are exponentially damped away from the inhomogeneities, radial structures are often observed. An example is the stable inhomogeneous state (see first snapshot of Fig. 4 in the main text). We have checked that this state is always obtained by direct simulations of the deterministic equation starting from various initial conditions (not shown). At this stage, it is unclear whether annular-like critical points with Morse index $\geq 1$ do exist or not. We show a typical scenario in Fig. 4 where one is able to identify a pair of radial solutions but which do not seem to be some zeros of $\nabla F_{\mathrm{aGL}}$ and saturate to a finite but small value of the residual no matter the NN size and hyperparameter changes.

## V. DISCUSSION ON PATHS SELECTION AND LINEAR STABILITY ANALYSIS

Our numerical analysis suggests that above some critical value of $\lambda$, we find that the reaction path from $\phi_I$ to $\phi_H$ goes through the critical states with the *highest number* of unstable directions. More precisely, when the critical states with $q$ bumps fit into the system, then, *either* (i) the reaction path goes through the critical states $\phi_{c,q}(x)$ (and displays also $q$ bumps), *or* (ii) the reaction path displays $q+1$ bumps, does not converge to $\phi_{c,q}(x)$ and crosses the separatrix elsewhere. Situation (i) corresponds to the parts of the curves in Fig. 3 (main text) where the action is locally increasing, while situation (ii) corresponds to the locally decreasing parts of the curves on the same plot. In other words, Fig. 3 (main text) shows that the reaction paths can display $q$ bumps *before* the corresponding critical states $\phi_{c,q}(x)$ emerges. The fact that the reaction paths go through the highest Morse index states is not observed for values of $\lambda < 0$ (when $\phi_0 > 0$).

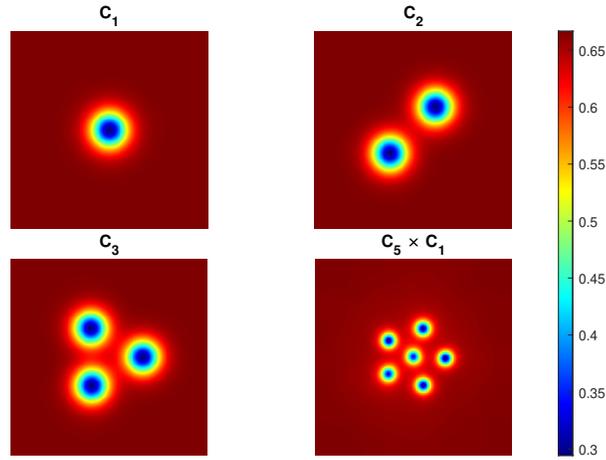

FIG. 3. Various spikes (critical solutions) obtained by minimizing (31) by deep neural networks and for $\phi_0 = 0.65$, $L = 44.7\nu^{1/2}$ and $\lambda = 2\nu$. The last one is for $L = 100\nu^{1/2}$.

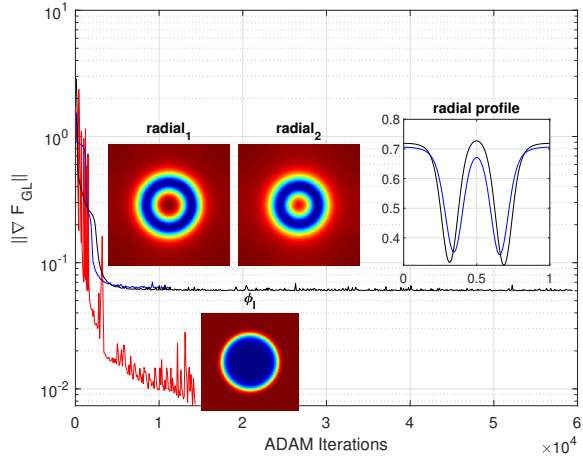

FIG. 4. An example of radial solutions which behave as pseudo-critical points for $\phi_0 = 0.65$, $L = 44.7\nu^{1/2}$ and $\lambda = 2\nu$. The inhomogeneous state is shown for comparison.

To gain insights on the selected reaction paths, we perform below a spectrum analysis of the operator acting on the perturbations around $\phi_I$. The analysis confirms that $\phi_I$ possesses stable direction only, and two marginally stable directions (Goldstone modes) corresponding to the mass conservation and to space translation invariance (due to the periodic boundary condition). Interestingly, the eigenvectors may display an oscillating profile reminiscent of the states along the instanton. However, the less stable eigenvector (corresponding to the less negative eigenvalue) does not correlate to the number of bumps selected along the instanton, as one could have expected.

The stability of the inhomogeneous state $\phi_I$ in 1-d is studied below as a function of the active term intensity. We consider here the spectral stability of $\phi_I$ giving the following eigenvalue problem:

$$\mathcal{M}\phi \equiv -\mathcal{A}''\phi = \sigma\phi \tag{32}$$

with $\mathcal{A}\phi = \nu\phi'' + (1 - 3\phi_I^2)\phi - 2\lambda\phi_I'\phi'$ and where $u'$ means $du/dx$. A first remark is that one can show that there is always two neutral modes denoted $\phi_1, \phi_2$ for the operator $\mathcal{M}$. By remarking that neutral modes for $\mathcal{A}$ are necessary neutral for $\mathcal{M}$ we observe that $\mathcal{A}\phi = 0$ has the nontrivial solution

$$\phi_1 = \phi_I'.$$

This is a direct consequence of $\phi_I$ solving $\nabla F_{\text{aGL}} = 0$. The other solution can be obtained using e.g. the Wronskian $W = \phi_1\phi_2' - \phi_2\phi_1'$ which gives $W = Ce^{2\alpha\phi_I}$. These neutral modes have a simple interpretation and correspond to Goldstone modes for the mass conservation and invariance by translation.

We proceed to calculate the leading spectrum of $\mathcal{L}$ as a function of the intensity $\alpha$ of the active term. The results are presented in Figure 5a, which shows the eigenvalues, and Figure 5b, which displays the corresponding eigenvectors. Our analysis reveals

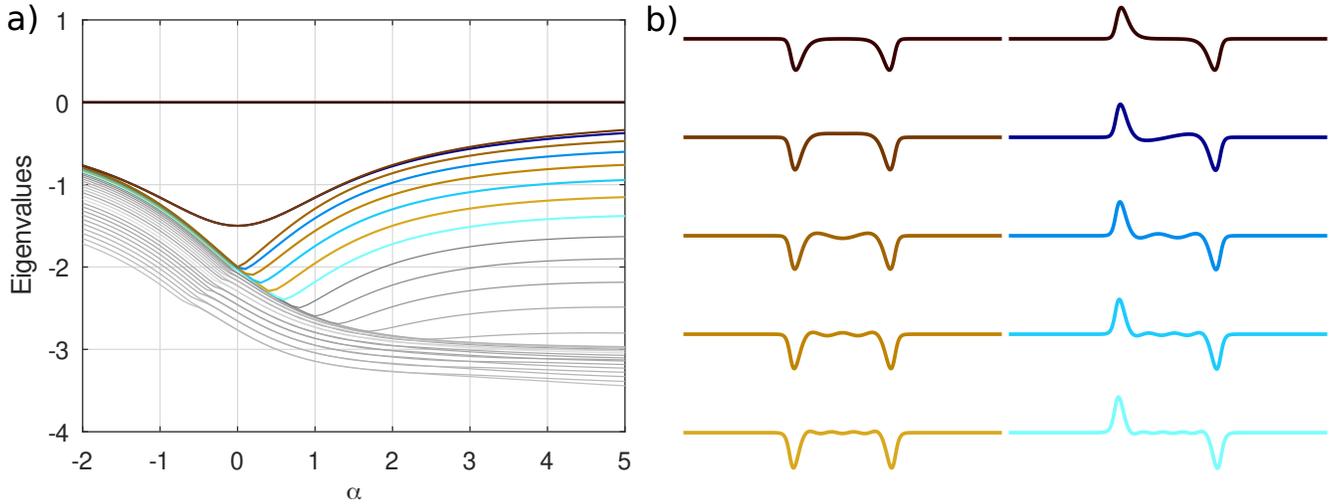

FIG. 5. Leading spectrum of $\mathcal{A}$ as a function of $\alpha \in (-2, 5)$ for $\nu = 10^{-4}$ and $\phi_0 = 0.65$. The 29 first eigenvalues are displayed on the left, the corresponding leading 10 eigenvectors (in color) are displayed on the right. The more the eigenvalue is negative, the more its eigenvector display bumps. The eigenvectors to the right are computed for $\alpha = 5$ with eigenvalues $(1, 2) : \sigma = (0, 0)$, $(3, 4) : \sigma = (-0.33, -0.37)$, $(5, 6) : \sigma = (-0.47, -0.60)$, $(7, 8) : \sigma = (-0.75, -0.94)$, $(9, 10) : \sigma = (-1.15, -1.38)$.

the emergence of "microstructures" or "bumps" at different wavenumbers, localized on the negative plateau of the ground state $\phi_I$. As $\alpha$ increases, higher wavenumber modes begin to activate from the background spectrum at specific values of $\alpha$. We also note that high wavenumber modes are more stable than small wavenumber modes. There is a well-defined spectral gap visible in Fig. 5a (blue curve) with corresponding eigenmodes 3 and 4 in Fig. 5b. This spectral gap decreases as $\alpha$ approaches infinity.

Our results help to explain why microstructures are observed in the minimum action paths. Although the inhomogeneous state is stable, noise excitations would tend to involve some of these modes. However, it is not clear why higher wavenumber bumps are preferred for the minimum action paths. Direct gMAM computations are necessary to reveal this behavior. Specifically, for this value of $\nu$, we would expect the one-bump regime to be involved in the minimum action paths, but this is not what is observed.

[1] R. Wittkowski, A. Tiribocchi, J. Stenhammar, R. J. Allen, D. Marenduzzo, and M. E. Cates, Nature Communications **5**, 4351 (2014).
[2] D. Donnelly and E. Rogers, American Journal of Physics **73**, 938 (2005).
[3] E. Forest and R. D. Ruth, Physica D: Nonlinear Phenomena **43**, 105 (1990).
[4] E. Vanden-Eijnden and M. Heymann, The Journal of Chemical Physics **128**, 061103 (2008).
[5] W. E, W. Ren, and E. Vanden-Eijnden, Phys. Rev. B **66**, 052301 (2002).
[6] W. E, W. Ren, and E. Vanden-Eijnden, Communications on pure and applied mathematics **57**, 637 (2004).
[7] R. Zakine and E. Vanden-Eijnden, Phys. Rev. X **13**, 041044 (2023).
[8] E. Simonnet, Journal of Computational Physics **491**, 112349 (2023).
[9] Z. Liu, W. Cai, and J. Z.-Q. Xu, Communications in Computational Physics **28**, 1970 (2020).
[10] D. Kingma and J. Ba, arXiv:1412.6980 (2014).
[11] P. W. Bates and G. Fusco, Journal of Differential Equations **160**, 283 (2000).



\end{document}